\documentclass[lettersize,journal]{IEEEtran}
\usepackage{cite}
\usepackage{hyperref}
\usepackage{amsmath,amssymb,amsfonts}
\usepackage{algorithmic}
\usepackage{graphicx}
\usepackage{textcomp}
\usepackage{xcolor}
\usepackage[font=small]{caption}
\usepackage{subcaption}
\usepackage{array}%
\usepackage{multirow}
\usepackage{algorithm}

\usepackage{url}
\begin{document}
\title{Diffusion-Aided Joint Source Channel Coding For High Realism Wireless Image Transmission}

\author{Mingyu~Yang,
        Bowen~Liu,
        Boyang~Wang, 
        and~Hun-Seok~Kim,~\IEEEmembership{Senior Member,~IEEE}% <-this % stops a space
\thanks{This work was funded in part by NSF CAREER \#1942806.} \thanks{M. Yang, B. Liu, B. Wang, and H. Kim are with the Department
of Electrical and Computer Engineering, University of Michigan, Ann Arbor,
MI, 48109 USA e-mail: (mingyuy@umich.edu; bowenliu@umich.edu; boyangwa@umich.edu; hunseok@umich.edu) }

}% <-this % stops a space

\maketitle

\newcommand{\figFront}{
\begin{figure*}[t]
\centering
\includegraphics[width=0.95\linewidth]{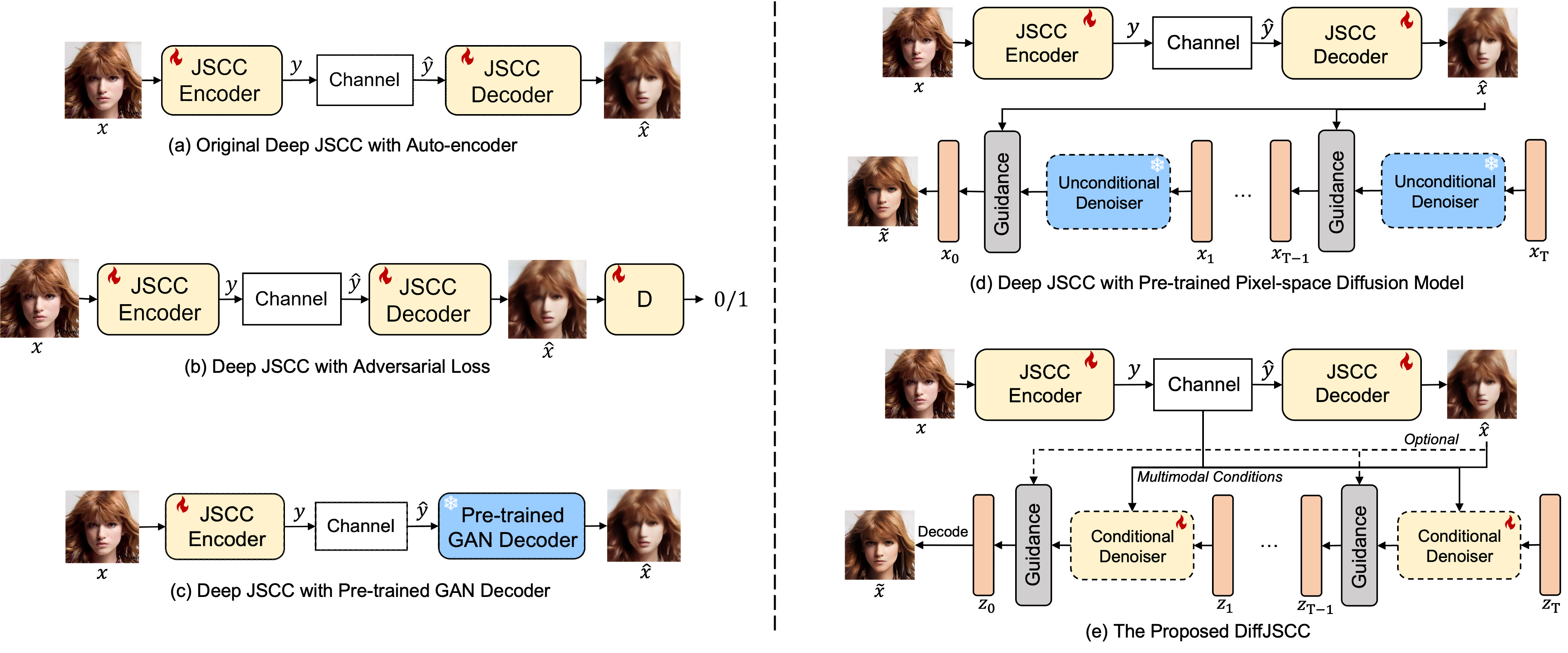}
\caption{Overall framework of the proposed DiffJSCC (e) and comparison with other existing deep JSCC structures (a-d).}
\label{fig:front}
\vspace{-0.2in}
\end{figure*}
}

\newcommand{\figAblation}{
\begin{figure}[t]
\centering
\includegraphics[width=\columnwidth]{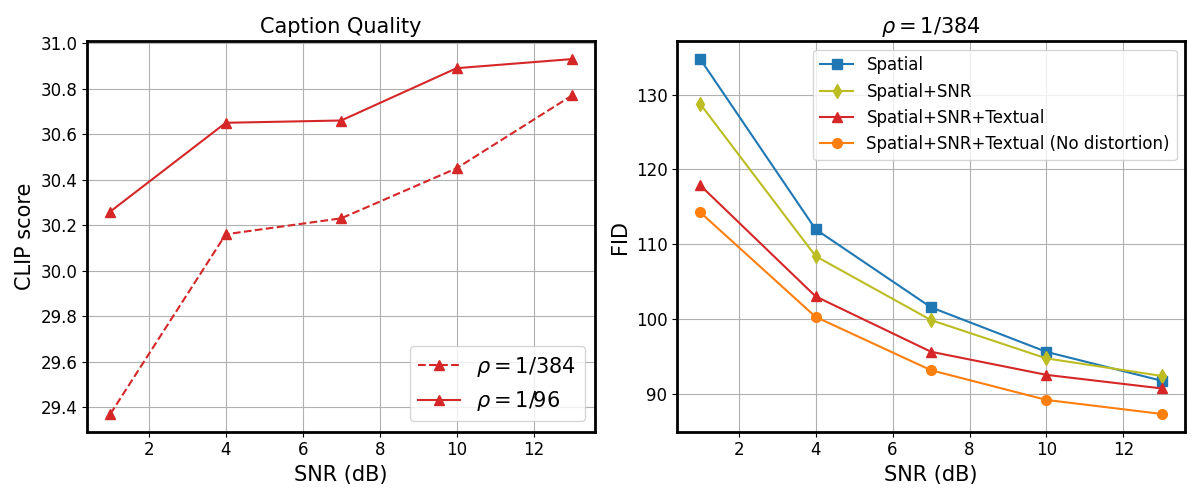}
\caption{Left: Analysis of the quality of the caption generated by BLIPv2 on the initial JSCC reconstruction. Right: Ablation experiments under various conditions for the control module.}
\label{fig:ablation}
\vspace{-0.2in}
\end{figure}
}

\newcommand{\figAblationVis}{
  \begin{figure*}[t]
    \centering
    \includegraphics[width=0.95\linewidth]{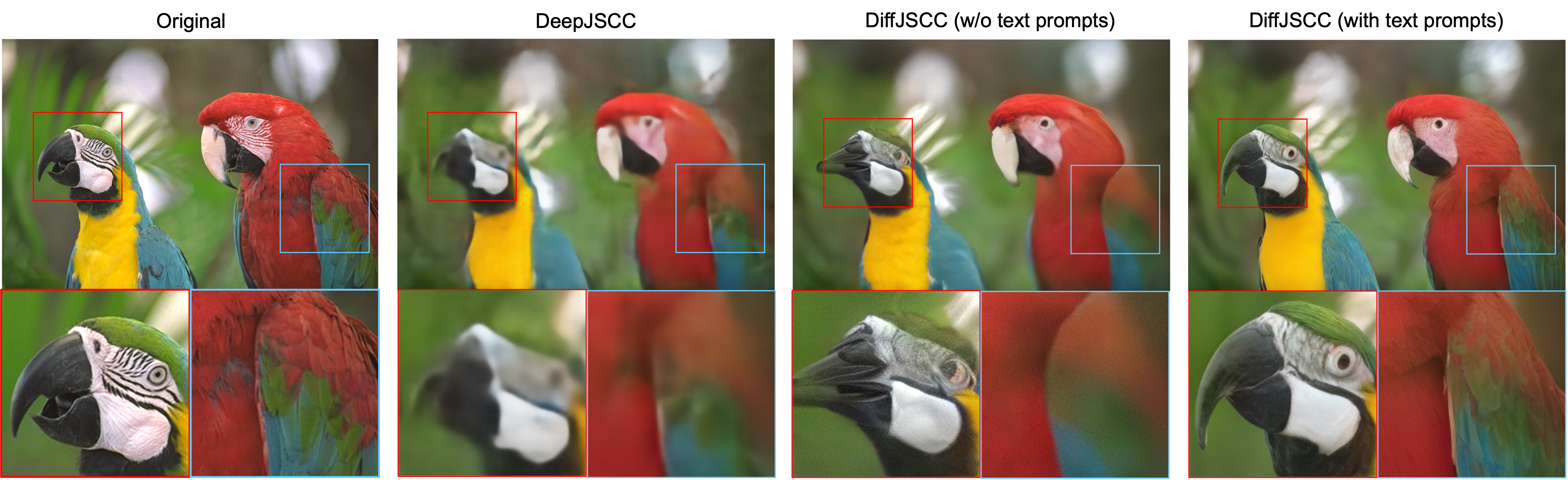}
    \caption{Example of generated image captions and the effect of them for the
final image reconstruction  }
    \label{fig:ablationVis}
    \vspace{-0.2in}
  \end{figure*}
}

\newcommand{\figStep}{
  \begin{figure}[t]
    \centering
    \includegraphics[width=\columnwidth]{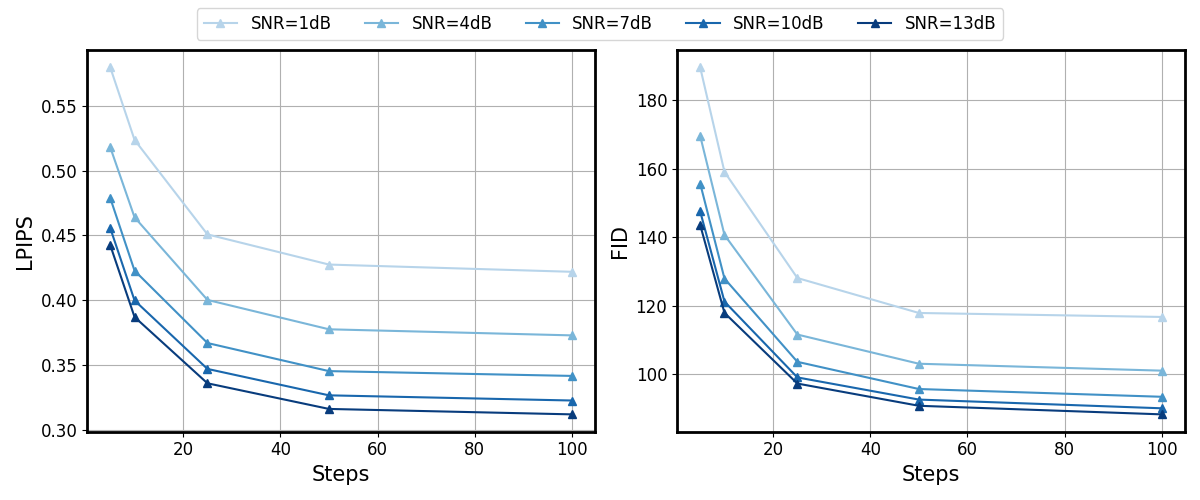}
    \caption{Effect of DDPM sampling steps with $\rho=1/384$. }
    \label{fig:ablationSteps}
    \vspace{-0.2in}
  \end{figure}
}

\newcommand{\figVisCelebA}{
  \begin{figure*}
    \centering
    \includegraphics[width=0.95\linewidth]{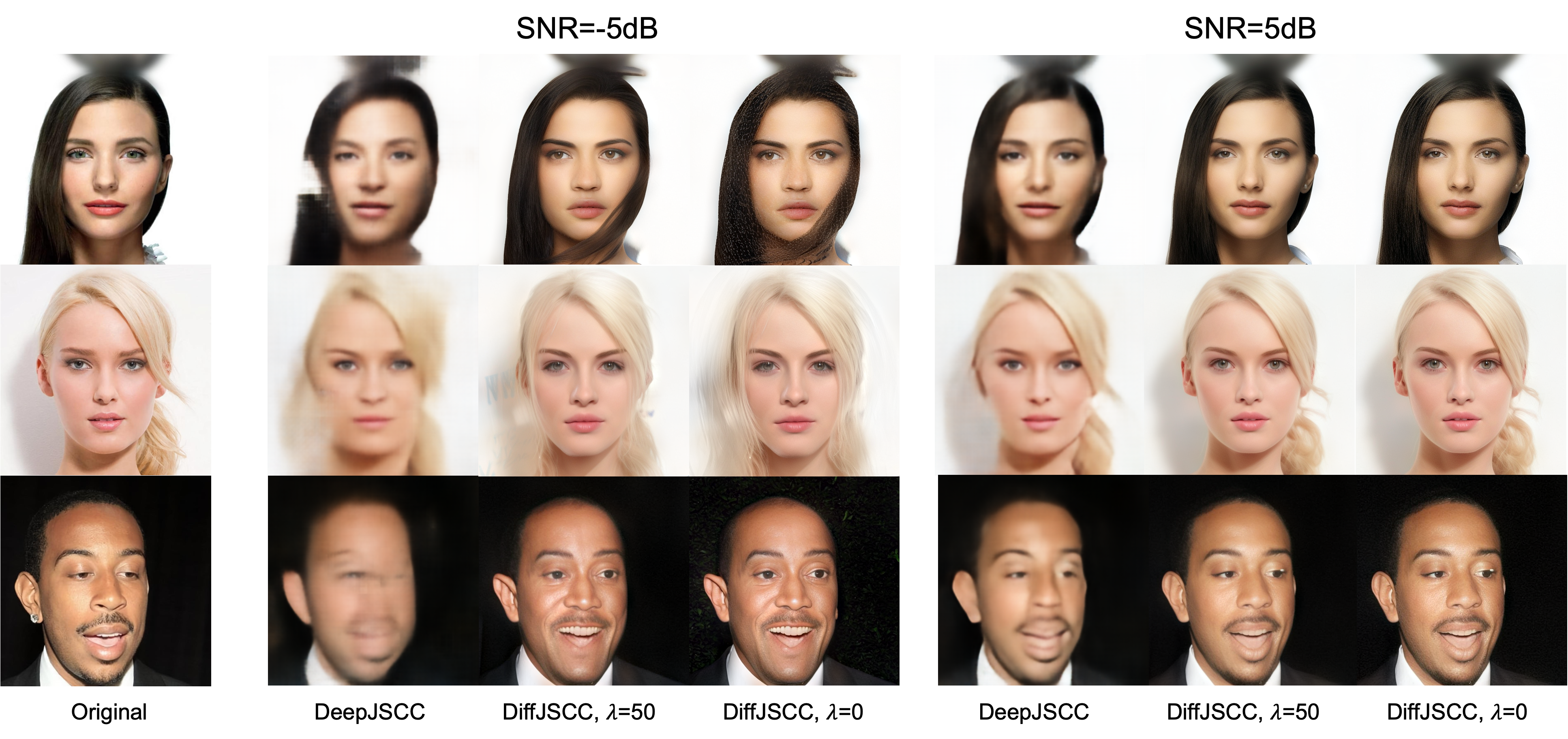}
    \caption{Visualizations on CelebAHQ with $\rho=1/768$ in both high and low SNR scenarios under the AWGN channel.}
    \label{fig:CelebAVis}
    %\vspace{-0.2in}
  \end{figure*}
}

\newcommand{\figVisADE}{
  \begin{figure}
    \centering
    \includegraphics[width=\columnwidth]{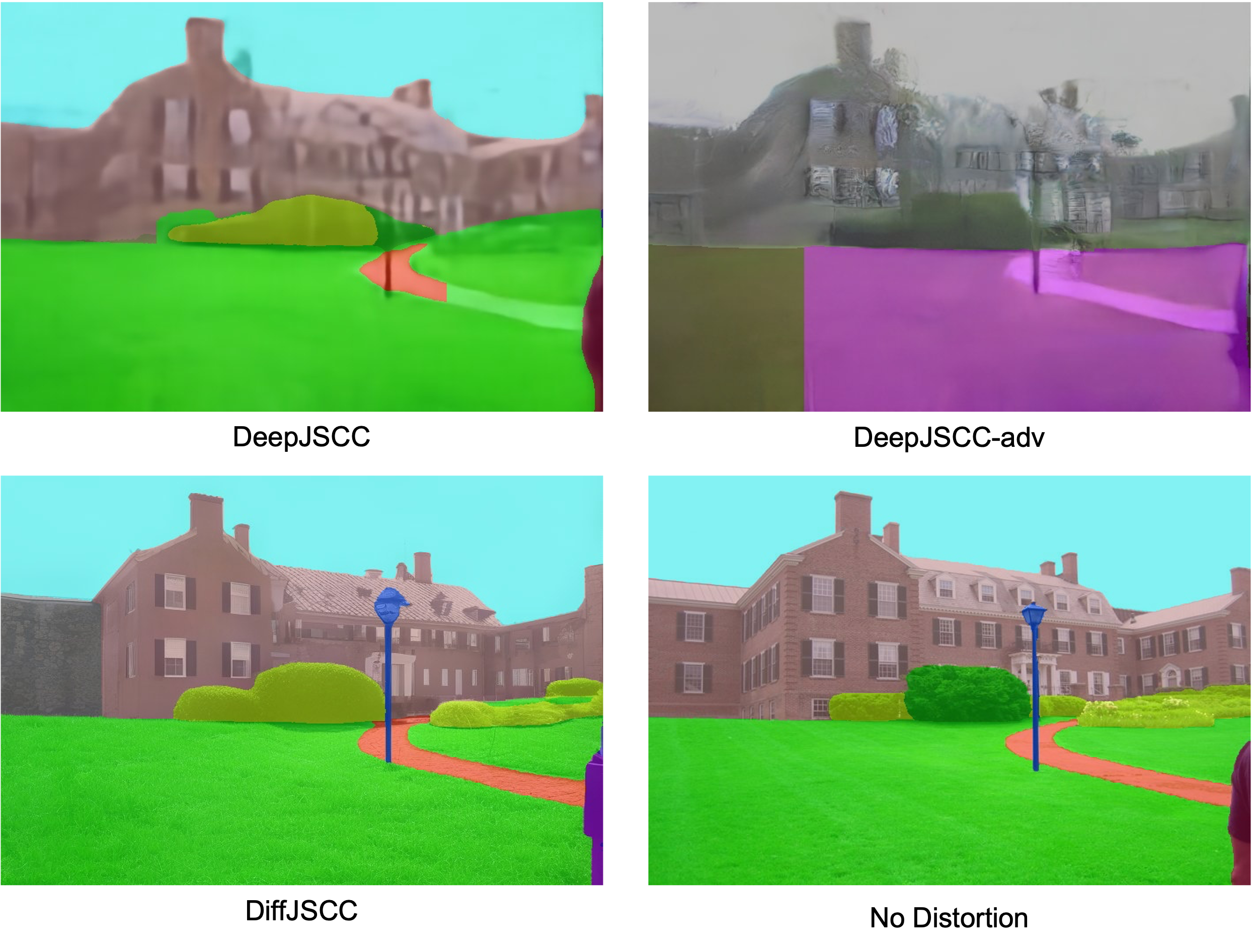}
    \caption{Visualizations on ADK20K with $\rho=1/384$ and 1dB SNR under the AWGN channel.}
    \label{fig:ADEVis}
    %\vspace{-0.1in}
  \end{figure}
}

\newcommand{\figControl}{
  \begin{figure}
    \centering
    \includegraphics[width=\columnwidth]{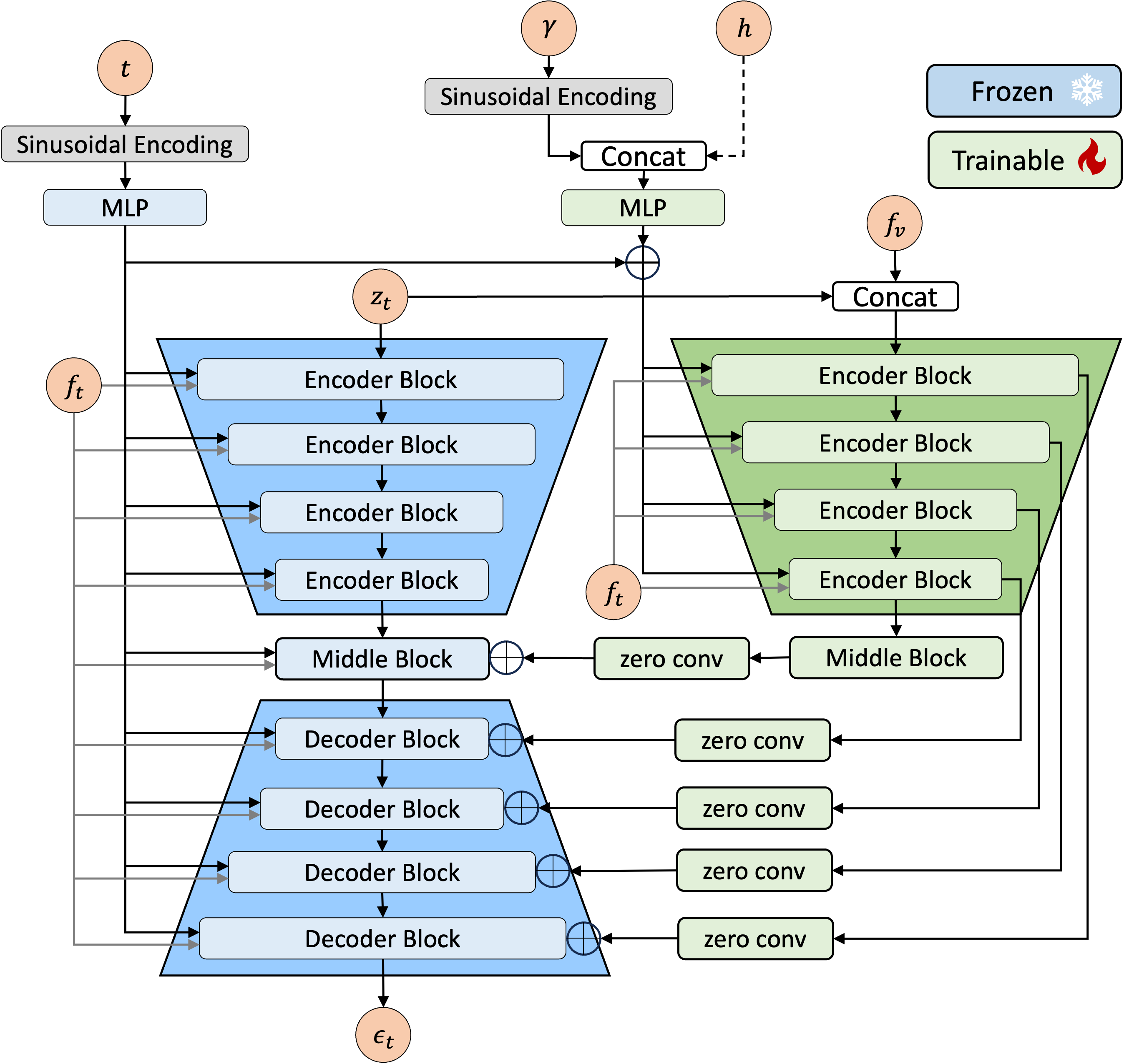}
    \caption{Network structure of the proposed denoiser $\epsilon_{\theta}$ in the conditional latent diffusion model. The pre-trained Stable Diffusion model is shown in blue and the fine-tuned control module is shown in green.}
    \label{fig:control}
    %\vspace{-0.2in}
  \end{figure}
}

\newcommand{\figStructure}{
  \begin{figure*}
    %\vspace{.2in}
    \centering
    \includegraphics[width=0.95\linewidth]{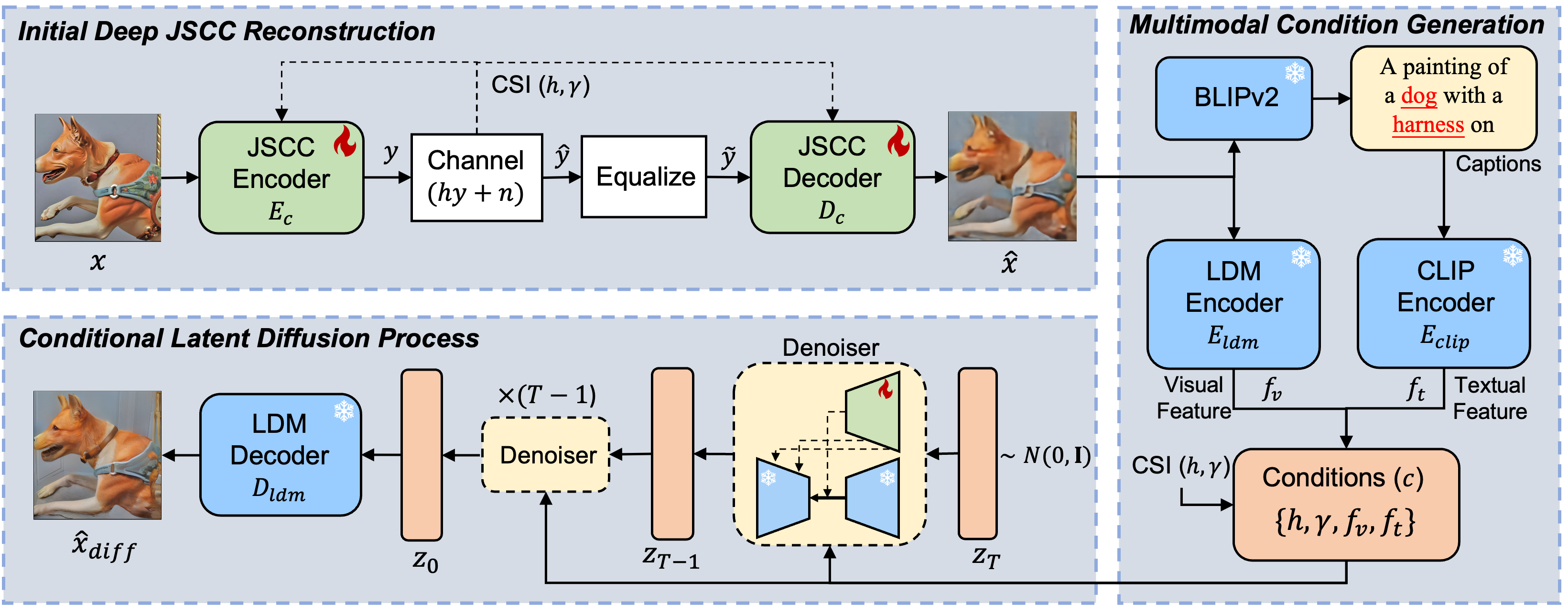}
    \caption{The general framework of the proposed DiffJSCC. In the transmitter, the image $x$ is encoded by a JSCC encoder to yield transmission symbols $y$. In the receiver, a preliminary image reconstruction $\hat{x}$ is generated by a JSCC decoder, which is used to produce multimodal (visual and text) conditions. The final step is the conditional latent diffusion process, which uses these multimodal conditions in the conditional denoiser to guide the image generation procedure. } 
    \label{fig:structure}
  \end{figure*}
}

\newcommand{\figKodakSNR}{
  \begin{figure*}
    %\vspace{.2in}
    \centering
    \includegraphics[width=\linewidth]{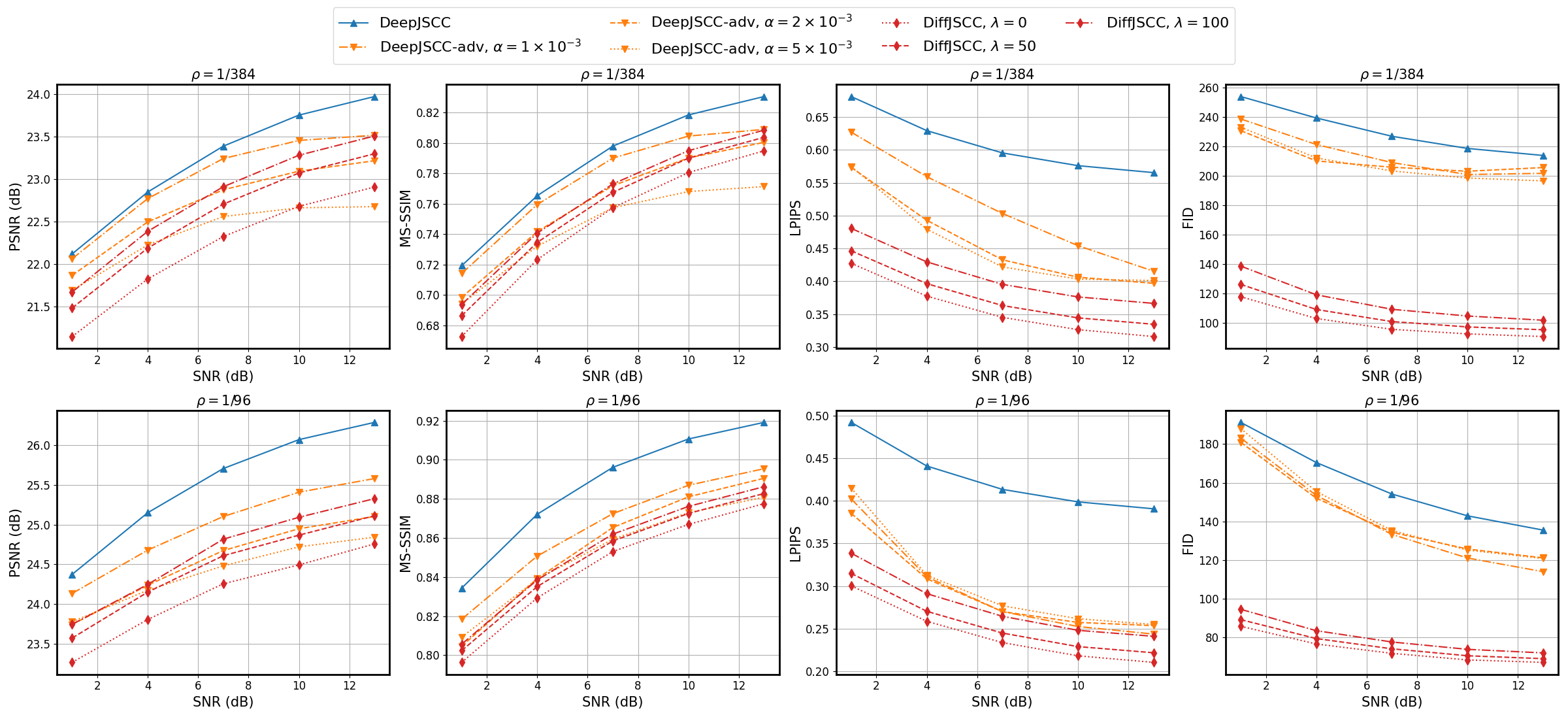}
    %\vspace{-0.1in}
    \caption{Evaluation of the proposed approach versus baselines across different SNR levels with transmission rates $\rho$ of $1/384$ (top) and $1/96$ (bottom) using the Kodak dataset under the AWGN channel. Lower LPIPS/FID scores indicate higher perceptual quality.}
    \label{fig:KodakSNR}
    \vspace{-0.1in}
  \end{figure*}
}

\newcommand{\figKodakSNRRAY}{
  \begin{figure*}
    %\vspace{.2in}
    \centering
    \includegraphics[width=\linewidth]{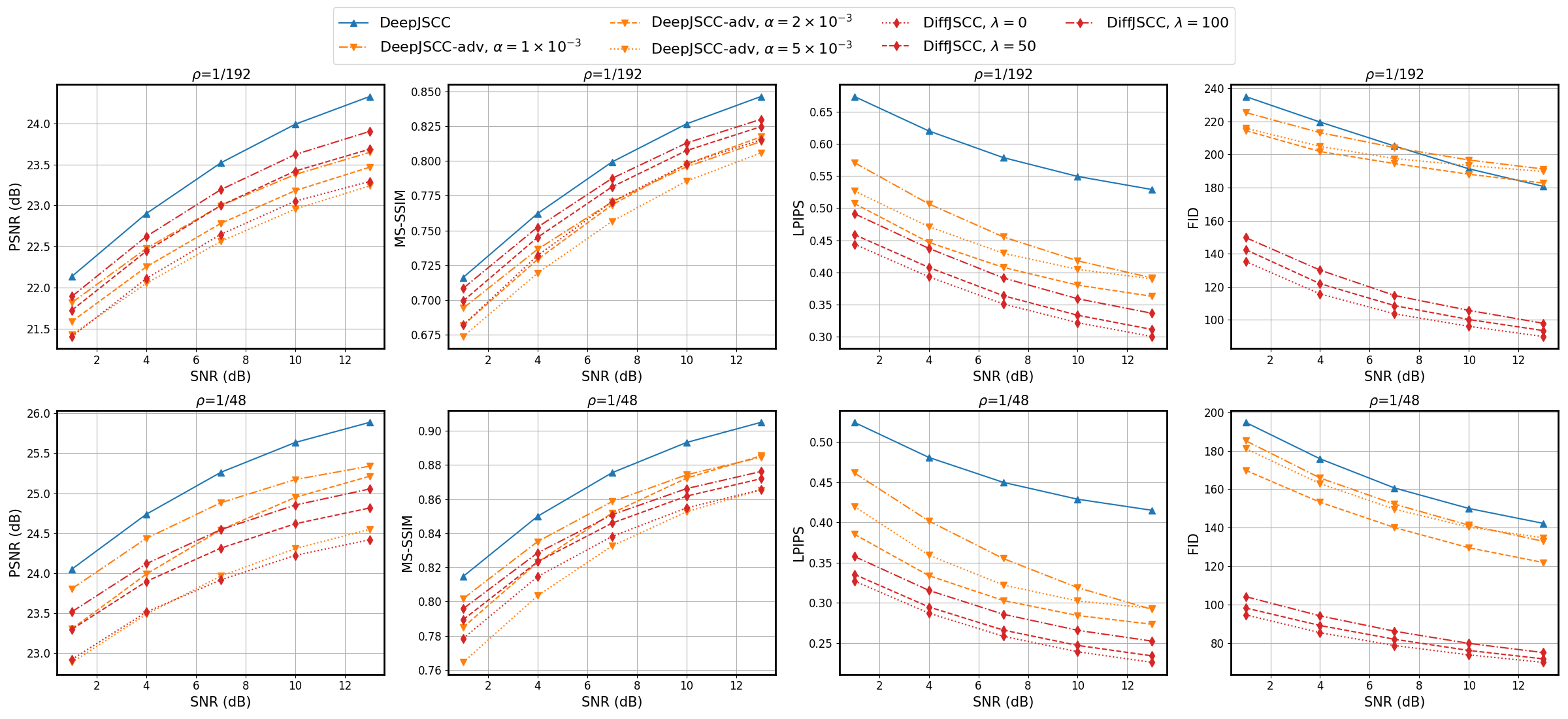}
    %\vspace{-0.1in}
    \caption{Evaluation of the proposed approach versus baselines across different SNR levels with transmission rates $\rho$ of $1/192$ (top) and $1/48$ (bottom) using the Kodak dataset under the Rayleigh Fading channel. Lower LPIPS/FID scores indicate higher perceptual quality.}
    \label{fig:KodakSNRRAY}
    \vspace{-0.1in}
  \end{figure*}
}

\newcommand{\figKodakRate}{
  \begin{figure*}[th]
    %\vspace{.2in}
    \centering
    \includegraphics[width=\linewidth]{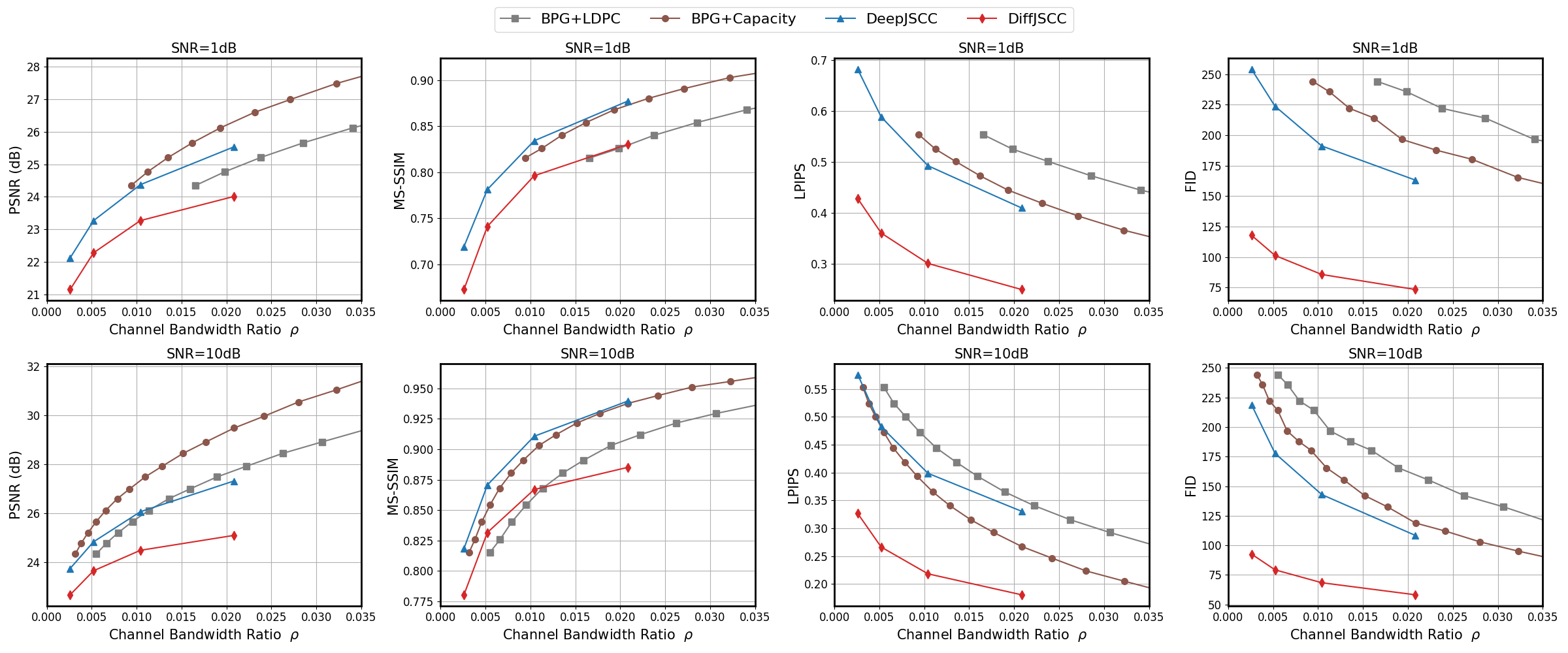}
    %\vspace{-0.1in}
    \caption{Evaluation of the proposed approach versus baselines across  different transmission rates under 1dB (top) and 10dB SNR (bottom) on the Kodak dataset under AWGN channel. Lower LPIPS/FID scores indicate higher perceptual quality.}
    \label{fig:KodakRate}
    \vspace{-0.1in}
  \end{figure*}
}

\newcommand{\figCelebA}{
  \begin{figure*}
    %\vspace{.2in}
    \centering
    \includegraphics[width=\linewidth]{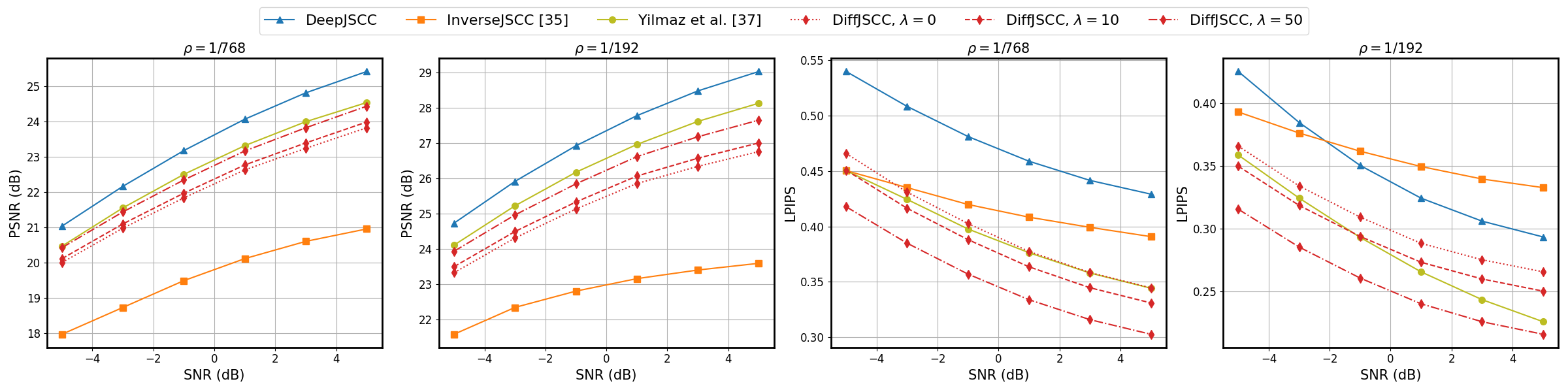}
    %\vspace{-0.1in}
    \caption{Comparison between the proposed method and baselines at various SNRs on CelebAHQ under the AWGN channel.}
    \label{fig:CelebASNR}
    %\vspace{-0.1in}
  \end{figure*}
}

\newcommand{\figVisual}{
  \begin{figure*}
    %\vspace{.2in}
    \centering
    \includegraphics[width=\linewidth]{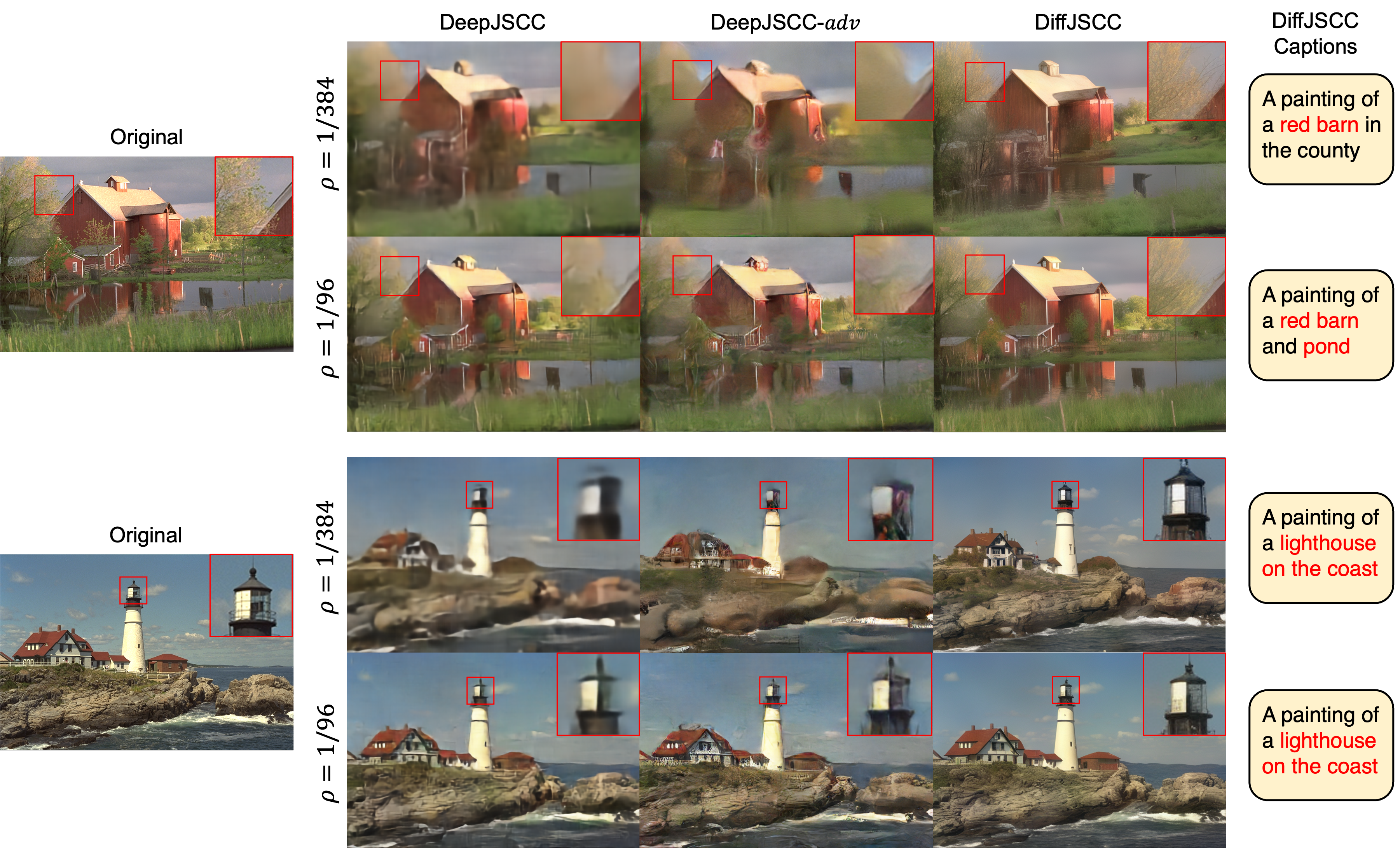}
    %\vspace{-0.1in}
    \caption{Visual comparison between DiffJSCC and baselines on the Kodak dataset at 1dB SNR under the AWGN channel. For DiffJSCC, we show both the final reconstructed images and the captions generated from the initial reconstruction. The keywords are marked in red. Zoom in for better perception.}
    \label{fig:KodakVis}
    \vspace{-0.1in}
  \end{figure*}
}

\newcommand{\figADE}{
  \begin{figure}[t]
    \centering
    \includegraphics[width=\linewidth]{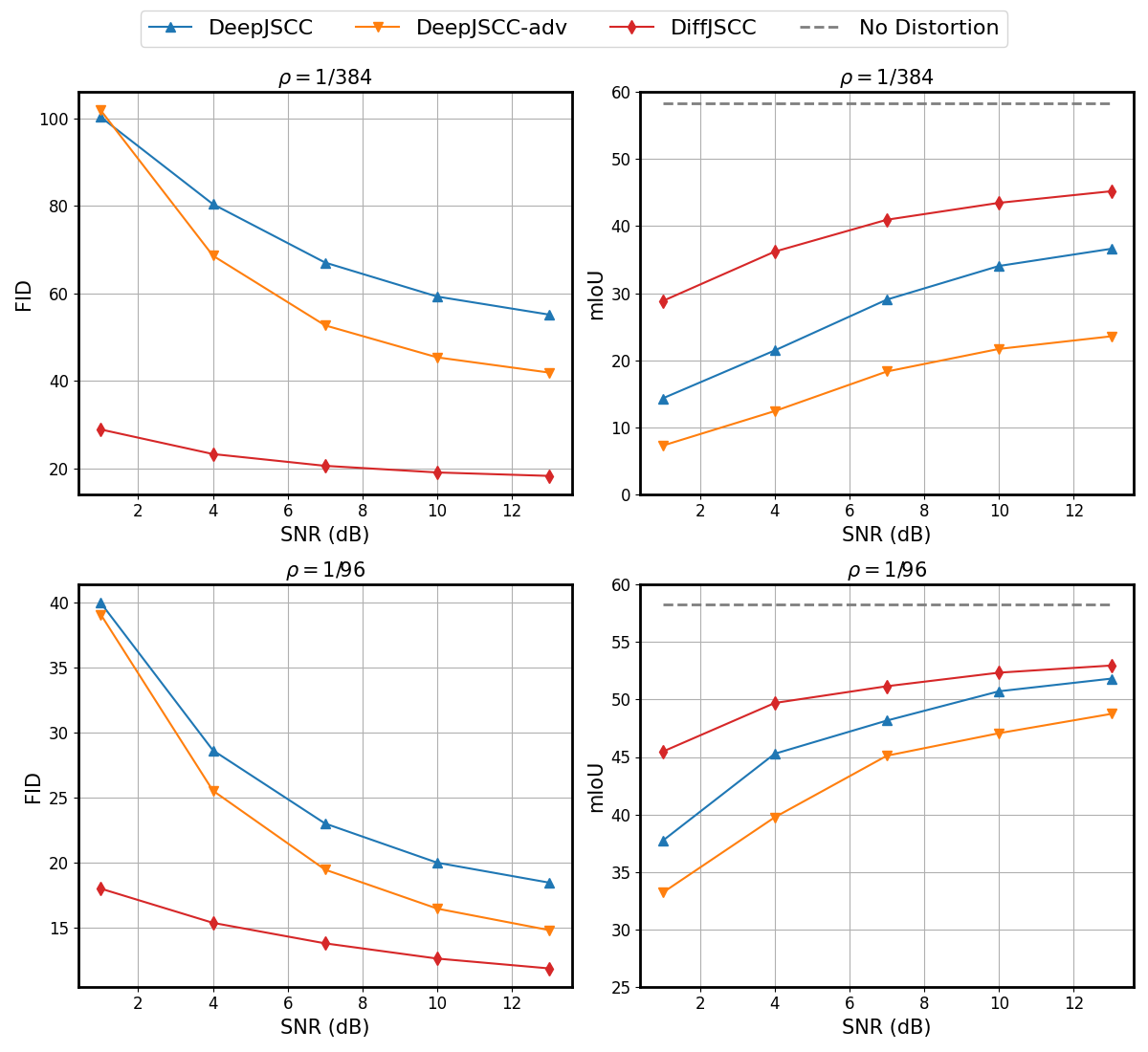}
    \caption{Results on the ADE20k dataset with $\rho=1/384$ and $\rho=1/96$. Higher mIoU scores indicate better segmentation.}
    \label{fig:figADE}
  \end{figure}
}

\newcommand{\algo}{
\begin{algorithm}[t]
\caption{Latent diffusion denoising process with intermediate latent guidance}\label{alg:alg1}
\begin{algorithmic}
\STATE \textbf{Input:} Time steps $T$, spatial features $f_{v}$, textual features $f_{t}$, channel SNR $\gamma$, channel gain $h$, scale factor $\lambda$, conditional denoiser $\epsilon_{\theta}$
\STATE \textbf{Output:} Generated image $\hat{x}_{diff}$
\STATE $\bar{c} \gets \{f_{v}, f_{t}, h, \gamma\}$
\STATE Sample $z_T$ from $\mathcal{N}(0, \mathbf{I})$
\STATE \textbf{for} $t$ from $T$ to $1$
\STATE \hspace{0.5cm}$ \tilde{z}_0 \gets \frac{z_t - \sqrt{1-\bar{\alpha}_t} \epsilon_{\theta}(z_t, \bar{c}, t)}{\sqrt{\bar{\alpha}_t}} $
\STATE \hspace{0.5cm}$ \hat{z}_0 \gets \tilde{z}_0 - \frac{\lambda}{C_lH_lW_l} (\tilde{z}_0 - f_{v}) $
\STATE \hspace{0.5cm}$ \mu(z_t, \hat{z}_0) \gets \frac{\sqrt{\bar{\alpha}_{t-1}}\beta_t}{1-\bar{\alpha}_t}\hat{z}_0 + \frac{\sqrt{1-\beta_t}(1-\bar{\alpha}_{t-1})}{1-\bar{\alpha}_t}z_t $
\STATE \hspace{0.5cm}Sample $z_{t-1}$ from $\mathcal{N}(\mu(z_t, \hat{z}_0), \sigma_t\mathbf{I})$
\STATE \textbf{end for}
\STATE $\hat{x}_{diff} \gets D_{ldm}(z_0)$
\STATE \textbf{return}  $\hat{x}_{diff}$
\end{algorithmic}
\label{alg1}
\end{algorithm}
}

\begin{abstract}
Deep learning-based joint source-channel coding (deep JSCC) has been demonstrated to be an effective approach for wireless image transmission. However, many current approaches utilize an autoencoder framework to optimize conventional metrics such as Mean Squared Error (MSE) and Structural Similarity Index (SSIM), which are inadequate for preserving the perceptual quality of reconstructed images. Such an issue is more prominent under stringent bandwidth constraints or low signal-to-noise ratio (SNR) conditions. To tackle this challenge, we propose DiffJSCC, a novel framework that leverages the prior knowledge of the pre-trained Stable Diffusion model to produce high-realism images via the conditional diffusion denoising process. First, our DiffJSCC employs an autoencoder structure similar to prior deep JSCC works to generate an initial image reconstruction from the noisy channel symbols. This preliminary reconstruction serves as an intermediate step where robust multimodal spatial and textual features are extracted. In the following diffusion step, DiffJSCC uses the derived multimodal features, together with channel state information such as the signal-to-noise ratio (SNR) and channel gain. These serve as conditions to guide the diffusion denoising process through a novel control module, which adjusts the output according to the multimodal conditions. To maintain the balance between realism and fidelity, an optional intermediate guidance approach using the initial image reconstruction is implemented. Extensive experiments on diverse datasets reveal that our method significantly surpasses prior deep JSCC approaches on both perceptual metrics and downstream task performance, showcasing its ability to preserve the semantics of the original transmitted images. Notably, DiffJSCC can achieve highly realistic reconstructions for 768$\times$512 pixel Kodak images with only 3072 symbols ($<$0.008 symbols per pixel) under 1dB SNR channels. The source code is available at \url{https://github.com/mingyuyng/DiffJSCC}.
\end{abstract}


\section{Introduction}

\IEEEPARstart{R}{obust} wireless image and video transmission have become crucial for emerging applications such as augmented reality (AR), virtual reality (VR), and autonomous driving, which commonly offload computation to remote servers. Conventional wireless image transmission utilizes a two-step scheme that first applies compression using codecs like Joint Photographic Experts Group (JPEG) or Better Portable Graphics (BPG), followed by a separate channel coding scheme such as a low-density parity-check code (LDPC) or Polar code for error protection. Nonetheless, these standard compression algorithms are prone to any bit errors and susceptible to dramatic quality degradation (so-called ``cliff effect'') when channel coding fails to correct codeword errors entirely. To address this challenge, joint source-channel coding (JSCC) has been proposed to improve error resilience and maintain image integrity during wireless transmission \cite{bozantzis2000combined, goodman1988using}.

\figFront

Recent advancements in Deep Neural Networks (DNNs) have significantly impacted fields including computer vision \cite{shoouri2023efficient, wang2024apisr, yang2022efficient}, natural language processing \cite{cho2014learning, vaswani2017attention}, and wireless communications \cite{ye2017power, yang2019ilps, hsiao2021super}, demonstrating their power in addressing complex problems including JSCC. Bourtsoulatze et al. \cite{bourtsoulatze2019deep} have introduced a groundbreaking deep learning-based JSCC (deep JSCC) framework that unifies image compression and error correction through an autoencoder architecture. This holistic end-to-end design, shown in Figure \ref{fig:front}a, surpasses conventional separate source-channel coding transmission techniques, especially in challenging wireless environments characterized by the Additive White Gaussian Noise (AWGN) channel and the Rayleigh Fading channel.

The scope of deep JSCC has been considerably expanded by recent developments that tailor its capabilities to various communication scenarios. For example, Kurka et al. \cite{kurka2020deepjscc} and Wu et al. \cite{wu2024transformer} integrated feedback mechanisms into JSCC. Yang et al. \cite{yang2021deep} introduced a JSCC variant compatible with orthogonal frequency division multiplexing (OFDM), which is adept at managing multipath channel effects. Bian et al. \cite{bian2023space} have extended JSCC's reach to multiple-input multiple-output (MIMO) channels. Additionally, Lee et al. \cite{lee2023deep} refined image reconstruction with a pre-trained channel denoiser at the receiver. Innovative neural network architectures, such as the Transformer \cite{vaswani2017attention} and the recently proposed Mamba \cite{gu2023mamba}, have also been adapted for JSCC \cite{yang2023witt, yang2023swinjscc, wu2024mambajscc}, promising further advancements.

Simultaneously, research has advanced techniques to enhance deep JSCC's adaptability to varying channel conditions. Xu et al. developed ADJSCC \cite{xu2021wireless}, which employs a Squeeze-Excitation structure \cite{hu2018squeeze} to adaptively respond to signal-to-noise ratio (SNR) variations. Yang et al. \cite{yang2022deep} introduced a policy network designed to modulate transmission rates in accordance with various channel statuses and image contents.  Dai et al. \cite{dai2022nonlinear} optimized rate adjustment using a learned entropy model, and Zhang et al. \cite{zhang2023predictive} leveraged a predictive oracle network for informed rate selection.  

Although these prior innovations mark significant progress, their focus on minimizing image distortion through metrics such as Peak Signal-to-Noise Ratio (PSNR) and Structural Similarity Index Measure (SSIM) does not necessarily translate to maintaining perceptual quality, particularly under bandwidth constraints or adverse channel conditions where images may lose texture and structural clarity.
Image compression and restoration studies have long faced the issue of blurry reconstructions. Blau et al. \cite{blau2018perception} highlighted the perception-distortion trade-off, revealing that optimizing rate-distortion criteria alone does not guarantee the preservation of natural image characteristics. To close this gap, integrating adversarial loss into generative models has become prevalent, substantially improving the visual quality of compressed images \cite{agustsson2019generative, mentzer2020high}. With the advent of diffusion models \cite{ho2020denoising} that excel in producing realistic textures in image generation, various diffusion models are now being applied to enhance the quality of reconstructed images even under high compression scenarios, promising a new frontier in image compression \cite{careil2023towards, theis2022lossy}.

The application of generative deep learning models has recently been explored in the domain of deep JSCC. Building on prior research, Yang et al. \cite{yang2022ofdm} and Wang et al. \cite{wang2022perceptual} incorporated adversarial loss using PatchGAN \cite{isola2017image} into deep JSCC frameworks (shown in Figure \ref{fig:front}.b), aiming to improve visual quality. Despite their efforts, the performance was hindered by the difficulties inherent in training effective Generative Adversarial Networks (GANs) from scratch. In a different approach, Ecenaz et al. \cite{erdemir2023generative} leveraged the pre-trained StyleGAN-v2 generator \cite{karras2020analyzing} to enhance the perceptual quality of transmitted facial images through deep JSCC  (shown in Figure \ref{fig:front}.c). However, the reliance on traditional distortion metrics such as mean squared error (MSE) during optimization tempered the gains in practice.
Yilmaz et al. \cite{yilmaz2023high} recently pioneered the integration of deep JSCC with pre-trained pixel-space diffusion models based on the zero-shot image restoration framework DDNM \cite{wang2022zero} (as shown in Figure \ref{fig:front}.d), achieving notable improvements in distortion and perceptual quality compared to Ecenaz et al.'s results \cite{erdemir2023generative}. Their approach involves initially generating an image reconstruction using a conventional JSCC autoencoder. Subsequently, this reconstructed image is utilized to guide the denoising process of the unconditional diffusion model via Range-Null space decomposition. However, since the diffusion process is regulated solely by the intermediate guidance, this method struggles to achieve high-quality image reconstructions, especially when the initial image reconstruction undergoes significant distortion under challenging channel conditions. Additionally, such a zero-shot generation process only supports images as the unimodal condition. 

%%%%%% previous paragraph %%%%%%%%%
%To address the dataset/domain mismatch issue, Zhang et al. recently have proposed ControlNet \cite{zhang2023adding}, a method for refining pre-trained diffusion models including Stable Diffusion \cite{rombach2022high}, to perform a more robust diffusion process under mismatched conditions where the target image domain is not initially covered during training. Those conditions occur with specific edge shapes/definitions, object motions, and varied content types like depth-map images and cartoons. By fine-tuning rather than comprehensive retraining on new datasets, ControlNet effectively minimizes the disparity between the capabilities of pre-trained model and the requirements of specific target datasets. This approach has proven beneficial across multiple disciplines enhancing image and video generation \cite{huang2023composer, ju2023humansd}, image restoration \cite{lin2023diffbir}, and image compression \cite{lei2023text+, careil2023towards}, thereby showcasing its versatility in improving the performance of diffusion-based generative models.
%%%%%%%%%%%%%%%%%%%%

\figStructure

Recent progress in image restoration \cite{lin2023diffbir} and compression \cite{lei2023text+, careil2023towards} has shifted away from solely depending on intermediate guidance for harnessing the prior knowledge in pre-trained diffusion models. Instead, these advancements explore the direct learning of conditional diffusion models through approaches like ControlNet \cite{zhang2023adding} and T2I-adaptor \cite{mou2024t2i}, demonstrating state-of-the-art performance and surpassing prior works such as DDNM\cite{wang2022zero}. These techniques involve fine-tuning the pre-trained diffusion model with additional lightweight control modules or adapters, and considerably reduce the dependence on intermediate guidance. In addition, these fine-tuning methods offer potential support for multimodal conditions to further enhance the control over the image generation process.

Leveraging the latest breakthroughs, we introduce DiffJSCC, a new framework devised to improve the perceptual quality of wirelessly transmitted images through conditional diffusion models as depicted in Figure \ref{fig:front}.e. Similar to Yilmaz et al's method \cite{yilmaz2023high}, our approach begins with initial image reconstruction via a standard autoencoder framework. In contrast, DiffJSCC focuses on directly learning a conditional diffusion model by fine-tuning the existing pre-trained diffusion model. On one hand, the distorted initial image reconstruction is treated as an input to the conditional denoiser instead of directly guiding the diffusion process, which ensures high-quality image generation even if the initial image reconstruction is highly distorted. On the other hand, this fine-tuning approach allows for multimodal conditions beyond the reconstructed image. Within DiffJSCC, we include text and channel state information (CSI) as additional conditions alongside the visual condition. Both visual and textual conditions are derived via pre-trained neural networks, obviating the need to develop new neural architectures and simplifying the training process. To fuse these multimodal conditions, DiffJSCC utilizes a controller module similar to that in Lin et al. \cite{lin2023diffbir}. The benefit of incorporating supplementary textual and CSI conditions has been demonstrated through our experiments. Instead of using pixel-domain diffusion models as in \cite{yilmaz2023high}, DiffJSCC adopts the advanced latent diffusion model known as Stable Diffusion \cite{rombach2022high}. The denoising procedure is aligned with the original denoising diffusion probabilistic model (DDPM) \cite{ho2020denoising}. Since DiffJSCC directly learns a conditional diffusion model, intermediate guidance is no longer necessary. However, we demonstrate that this mild intermediate guidance remains beneficial for balancing image realism with fidelity. Empirical evaluations across diverse image datasets reveal that DiffJSCC surpasses traditional separate source-channel coding schemes and prior deep JSCC benchmarks, showing superior results in human-perceptual quality and practical task efficacy, particularly in scenarios of restricted bandwidth and challenging channel conditions.

We summarize our primary contributions as follows:
\begin{itemize}
  \item \textit{Novel Diffusion-based JSCC Framework:} We present a novel JSCC framework that utilizes a pre-trained text-to-image Stable Diffusion model to produce realistic image reconstructions. We develop a custom control module to adapt the Stable Diffusion model for conditional generation for particular datasets. To the best of our knowledge, this is the first instance of integrating the Stable Diffusion model into deep JSCC.
  
  \item \textit{Enhanced Multimodal Conditioning:} Our framework uniquely extracts multimodal conditions to better guide the Stable Diffusion model by including visual and textual features and the channel information. We demonstrate that these diverse conditions substantially improve the performance.
  
  \item \textit{Balancing Fidelity and Realness:} We present a denoising guidance process that derives intermediate features with an initial JSCC reconstruction. This approach adeptly balances the image fidelity with perceptual realness.
  
  \item \textit{Flexible and Versatile Design:} DiffJSCC's model-agnostic design is compatible with various advanced deep JSCC architectures for initial image reconstruction. Besides, DiffJSCC omits any additional components at the transmitter, suiting mobile and IoT applications that offload heavy computations to remote servers.
  
  \item \textit{Improved Image Quality with Low Transmission Rates:} Our comprehensive experimentation with multiple datasets confirms that our approach significantly surpasses benchmarks in perceptual quality and task performance, particularly in bandwidth-constrained and noisy channel situations. Notably, DiffJSCC can achieve realistic reconstructions of $768\times512$ pixel images with fewer than $0.008$ symbols (channel uses) per pixel over a 1dB SNR AWGN channel. 
  
\end{itemize}

\section{Proposed Method}

The general architecture of our proposed DiffJSCC is illustrated in Figure \ref{fig:structure}, which contains three main components. First, we perform Initial Deep JSCC Reconstruction (Sec, \ref{JSCC}) using traditional autoencoder structures. Then, we perform Multimodal Condition Generation (Sec. \ref{RCG}) to extract multimodal spatial and textual features from the initial JSCC reconstruction. Lastly, the multimodal features, together with the channel state information, are treated as the conditions and fed to the Conditional Latent Diffusion Process (Sec. \ref{CLDM}) to generate the final refined image reconstruction. More details about the three components are introduced in the following sections.

\subsection{Initial Deep JSCC Reconstruction} \label{JSCC}

The input image is denoted as $x \in \mathbb{R}^{C \times H \times W}$, with $C$ indicating the number of color channels, and $H$ and $W$ representing the height and width, respectively. Within the transmitter, a JSCC encoder, $E_c$, encodes the image $x$ into a sequence of complex-valued channel input symbols $y \in \mathbb{C}^K$. These $K$ symbols are subsequently transmitted through a noisy communication channel while each symbol utilizes the channel once. The rate of the transmission is quantified as the channel bandwidth ratio (CFR), which is referred as $\rho$ and is defined as:
\begin{equation}
    \rho = \frac{K}{C\times H\times W}.
\end{equation}
To comply with power constraints, the final encoder layer normalizes the transmitted symbols $y$ to ensure an average power output of $\bar{P}$. Let $\tilde{y}$ represent the complex symbols prior to normalization; the normalization procedure is expressed mathematically as:
\begin{equation}
    y = \sqrt{K\bar{P}}\frac{\tilde{y}}{\sqrt{\tilde{y}'\tilde{y}}}.
\end{equation}

The wireless channel in our study is modeled as a conditional distribution, with the received symbols being drawn according to $\hat{y} \sim p(y|\varepsilon)$, where $\varepsilon$ encapsulates the channel parameters. In this work, we consider both the AWGN channel and Rayleigh Fading channel, and the channel transfer function can be represented as:
\begin{equation}
    \hat{y} = hy + n, \;\; n \sim \mathcal{CN}(0, \sigma^2 \mathbf{I}_K).
\end{equation}
Here, the noise $n \in \mathbb{C}^K$ is assumed to have a circularly symmetric complex Gaussian distribution with $\sigma^2$ symbolizing the noise's power level. $h \in \mathbb{C}$ denotes the channel gain. For AWGN channel, $h$ is set to 1. For the Rayleigh Fading channel, the channel gain $h$ is a random variable and it follows $h \sim  \mathcal{CN}(0, H)$. The channel's SNR, denoted as $\gamma$, is defined by the ratio $\gamma = \bar{P}\mathbb{E}[|h|^2] / \sigma^2$. In our experiments, we standardize the average power $\bar{P}$ to 1 for consistency. Following the prior works \cite{xu2021wireless, tung2022deepjscc}, both the transmitter and receiver are assumed to have perfect knowledge of the channel gain $h$ and SNR $\gamma$. Thus, at the receiver side, we first perform equalization to the received symbols $\hat{y}$, which is:
\begin{equation}
    \tilde{y} = \frac{h^*}{|h|^2}\hat{y},
\end{equation}
where $h^*$ denotes the complex conjugate of $h$. 

After that, a JSCC decoder ($D_c$) takes the equalized channel symbols $\tilde{y}$ as input and generate the initial reconstruction $\hat{x} \in \mathbb{R}^{C \times H \times W}$. 
In this paper, we employ a JSCC architecture akin to the one described in Yang et al.\cite{yang2022deep}. The encoder and decoder could take the CSI information $\gamma$ and $h$ and are trained jointly. The encoding and decoding processes are expressed as: 
\begin{equation} 
    y = E_c(x, h, \gamma), \; \; \hat{x} = D_c(\hat{y}, h, \gamma).
\end{equation} 
The aim is to minimize the MSE between the original image $x$ and the decoder's output $\hat{x}$, with the following loss function deployed for optimization: 
\begin{equation} \mathcal{L}_{JSCC} = \mathbb{E}_{x, h, n}\left[ \|x - \hat{x} \|_2^2 \right], 
\end{equation} 
A key aspect of our approach is its model-agnosticism, facilitating seamless integration with more advanced deep JSCC frameworks. While we demonstrate the effectiveness of DiffJSCC using one JSCC structure, the design is inherently adaptable to other more recent architectures such as SwinJSCC\cite{yang2023swinjscc} or NTSCC\cite{dai2022nonlinear}.

Rather than directly using the received symbols $\hat{y}$ as conditioning inputs for the conditional latent diffusion model, the initial JSCC reconstruction acts as an intermediate step and exhibits three primary benefits: First, it enables separate training of the JSCC encoder and the conditional denoiser, which allows more stable training compared to a joint training approach. Second, the initial reconstruction $\hat{x}$ harnesses the pre-trained LDM encoder to capture salient spatial features, thus avoiding the need to develop an encoder from scratch as originally proposed by the ControlNet framework\cite{zhang2023adding}. Last, the initial reconstruction $\hat{x}$ facilitates the extraction of supplementary textual descriptors using pre-trained image captioning models such as BLIPv2\cite{li2023blip} and the CLIP encoder\cite{radford2021learning}, enriching the input conditions for the diffusion model.

% Nevertheless, our proposed methodology is versatile and can be adapted with minimal effort to accommodate alternative channel models. We adhere to the established convention, as presented in preceding studies\cite{xu2021wireless, yang2022deep}, that the channel SNR is known a priori at both the transmitting and receiving ends. This assumption simplifies the model's architecture, enabling a single unified model to reliably handle a span of channel SNR conditions.

% At the receiver, image reconstruction is structured into two core stages: Robust Condition Generation (Section \ref{RCG}) and Conditional Latent Diffusion Process (Section \ref{CLDM}). Initially, during the Robust Condition Generation stage, resilient multimodal features are extracted from the noisy received channel symbols to accurately reflect characteristics of the original image and channel state. Subsequently, the Conditional Latent Diffusion Process takes these features as conditioning inputs to a latent diffusion model (LDM). Leveraging ControlNet's architecture \cite{zhang2023adding}, our framework adopts the pre-trained Stable Diffusion model augmented with a control module that fine-tunes the denoising operation for specific conditions, offering high-realism image reconstructions.

\subsection{Multimodal Condition Generation} \label{RCG}

Leveraging the initial reconstruction $\hat{x}$, our framework extracts multimodal features with the help of pre-trained neural networks, eliminating the need for additional training specific to our approach. The spatial features $f_{v}$ with a size of $4\times H/8 \times W/8$ are extracted by the image encoder $E_{ldm}$ from the pre-trained Stable Diffusion model:
\begin{equation}
   f_{v} = E_{ldm}(\hat{x}).
\end{equation}

In parallel, given that the Stable Diffusion model is inherently a text-to-image generative model, it is natural to leverage textual information as a supplementary condition. Noticing the presence of an initial image reconstruction, we propose to utilize the pre-trained image captioning model BLIPv2 \cite{li2023blip} denoted as $E_{blip}$ to generate textual descriptions for $\hat{x}$. Following the initial captioning step, these descriptions are further encoded by the pre-trained CLIP encoder $E_{clip}$\cite{radford2021learning} to obtain textual features $f_{t} \in \mathbb{R}^{77 \times 768}$:
\begin{equation}
   f_{t} = E_{clip}(E_{blip}(\hat{x})).
\end{equation}
An alternative approach to leveraging textual data is by deploying image captioning models at the transmitter end and sending the generated captions along with the image. Nevertheless, since image captioning models often involve significant computational requirements, this method could substantially increase the transmitter's processing load, making it impractical for mobile systems. It also incurs the overhead of sending text captions through the noisy channel. Within our method, despite the potential quality degradation in captions due to the distortion of initial image reconstructions, the inclusion of $f_{t}$ generated from $\hat{x}$ has been shown to be beneficial as it encapsulates content-related \textit{keywords}.

The CSI information, described by the channel SNR $\gamma$ and the channel gain $h$, plays a crucial role alongside the multimodal features $f_{v}$ and $f_{t}$ as it provides insight into the extent of distortion that initial image reconstruction experiences. After integrating the CSI, the set of conditions $\bar{c}$ utilized by our conditional diffusion model is formulated as:
\begin{equation}
    \bar{c} = \{ f_{v}, f_{t}, h, \gamma \}.
    \label{condition}
\end{equation}
These diverse conditions work together to strengthen the conditional diffusion process, improving the model's capability to reconstruct images with increased authenticity and accuracy, faithfully reconstructing the originally transmitted content.

\figControl
\subsection{Conditional Latent Diffusion Model} \label{CLDM}

The proposed DiffJSCC framework takes the set of conditions $\bar{c}$ as the input of the conditional latent diffusion process as depicted in Fig. \ref{fig:structure}. The diffusion process starts with a random input $z_{T}$ sampled from $\mathcal{N}(0, \mathbf{I})$ and iterates the denoising process with the multi-modal condition set $\bar{c}$ using the proposed denoiser (Fig. \ref{fig:control}) until the denoised output $z_{0}$ is obtained. The final output image $\hat{x}_{diff}$ is generated using an LDM decoder on $z_{0}$: $\hat{x}_{diff} = D_{ldm}(z_{0})$.

\subsubsection{Preliminary of Latent Diffusion Model}

Ho et al. introduced the Denoising Diffusion Probabilistic Model (DDPM), a generative approach that iteratively refines a sample from a noise distribution towards a target data distribution using a sequence of denoising operations \cite{ho2020denoising}. To improve this process for efficiency and training stability, Rombach et al. proposed the Stable Diffusion method in which a variational autoencoder (VAE) initially maps images into a lower-dimensional latent space, followed by diffusion denoising steps \cite{rombach2022high}. The forward diffusion process employs a Markov chain that incrementally introduces Gaussian noise to the latents at each time step $t$, with a variance denoted by $\beta_t \in (0, 1)$, applied to the state $z_{t-1}$ from the previous step, which is:
\begin{equation}
    z_t = \sqrt{1-\beta_t}z_{t-1} + \sqrt{\beta_t}\epsilon,
    \label{forward}
\end{equation}
where $\epsilon$ denotes the noise sampled from a standard Gaussian distribution $\mathcal{N}(0, \mathbf{I})$. Summarized over the time steps from $1$ to $t$, the forward process is represented as:
\begin{equation}
    z_t = \sqrt{\bar{\alpha}_t}z_0 + \sqrt{1-\bar{\alpha}_t}\epsilon.
    \label{forward}
\end{equation}
Here, $z_0 = E_{ldm}(x)$ is the latent representation of the original image encoded by the VAE, $\alpha_t = 1 - \beta_t$, and $\bar{\alpha}_t = \prod_{i=1}^t\alpha_i$ is the cumulative product of $\alpha_i$ from the initial time step to time step $t$, reflecting the progression of the diffusion process.

The aim of the denoising process is to estimate the prior latent state $z_{t-1}$ from the current state $z_t$. According to \cite{ho2020denoising}, the probability $p_{\theta}(z_{t-1}|z_{t})$ can be solved by minimizing the Evidence Lower Bound (ELBO), which is equivalent to minimizing the KL-divergence between $p_{\theta}(z_{t-1}|z_{t})$ and the posterior distribution $q(z_{t-1}|z_t, z_0)$. $q(z_{t-1}|z_t, z_0)$ can be modeled as a Gaussian $\mathcal{N}(z_{t-1};\mu_t(z_t, z_0), \sigma_t^2\mathbf{I})$. The mean of this distribution, $\mu_t(z_t, z_0)$, can be parameterized as: 
\begin{equation}
\begin{split}
    \mu_t(z_t, z_0) & = \frac{\sqrt{\bar{\alpha}_{t-1}}\beta_t}{1-\bar{\alpha}_t}z_0 + \frac{\sqrt{1-\beta_t}(1-\bar{\alpha}_{t-1})}{1-\bar{\alpha}_t}z_t\\
    & = \frac{1}{\sqrt{\alpha_t}}(z_t-\epsilon\frac{1-\alpha_t}{\sqrt{1-\bar{\alpha}_t}}).
\end{split}
\label{posterior}
\end{equation}
By approximating $p_{\theta}(z_{t-1}|z_{t})$ with a Gaussian distribution $\mathcal{N}(z_{t-1}; \mu_{\theta}(z_t, t), \sigma_t^2\mathbf{I})$, the KL-divergence is simplified as minimizing the difference between $\mu_t(z_t, z_0)$ and $\mu_{\theta}(z_t, t)$. Parameterizing $\mu_{\theta}(z_t, t)$ as
\begin{equation}
    \mu_{\theta}(z_t, t) = \frac{1}{\sqrt{\alpha_t}}(z_t-\epsilon_{\theta}(z_t, t)\frac{1-\alpha_t}{\sqrt{1-\bar{\alpha}_t}}),
    \label{mean}
\end{equation}
the problem is further simplified as accurately approximating the noise component $\epsilon$, which is facilitated by a neural network denoted as $\epsilon_{\theta}$. The network is typically conditioned on a context variable $c$ that can embed additional information such as text descriptions. Consequently, the network $\epsilon_{\theta}$ is trained to minimize the discrepancy between the true noise $\epsilon$ and its estimated value, using the loss function:
\begin{equation}
    \mathcal{L}_{ldm} = \mathbb{E}_{z_0, c, t, \epsilon}||\epsilon - \epsilon_{\theta}(\sqrt{\bar{\alpha}_t}z_0 + \sqrt{1-\bar{\alpha}_t}\epsilon, c, t)||_2^2.
    \label{ldm}
\end{equation}
This training paradigm ensures that over time, the network becomes proficient at deranging the noise from the corrupted data, steering the reverse-diffusion towards accurate data reconstruction.

\subsubsection{The Denoiser Structure}

%We propose a new denoiser $\epsilon_{\theta}$ structure as depicted in Figure \ref{fig:control}. It utilizes the UNet architecture \cite{ronneberger2015u} from the Stable Diffusion denoiser (shown in blue) which remains unchanged during training. 
The structure of the proposed denoiser, $\epsilon_{\theta}$, is illustrated in Figure \ref{fig:control}. The UNet architecture \cite{ronneberger2015u} of the Stable Diffusion denoiser is depicted in blue and remains unchanged during training. 
This architecture comprises of four encoder blocks, four decoder blocks, and one middle block. Each encoder/decoder block includes two main blocks and one convolutional block for scaling operations. The middle block contains two main blocks. Each main block consists of 4 ResNet and 2 Vision Transformers (ViTs). The ViTs integrate self-attention and cross-attention mechanisms to facilitate interaction with external prompts, such as text encoded by the CLIP text encoder.

%The green blocks denote the control module, which is inspired by ControlNet \cite{zhang2023adding}. This module employs the same UNet encoder and mid block configuration as Stable Diffusion, and it starts with the same pre-trained weights. Therefore, the control module comprises 12 encoding blocks along with 1 middle block. The 12 encoding blocks are set across 4 resolutions $(64 \times 64, 32 \times 32, 16 \times 16, 8 \times 8)$. The multi-scale outputs of the control module are then integrated into the 12 skip-connected layers and 1 middle block of the UNet from Stable Diffusion. Prior to this integration, the output of the control module will pass through several individual $1 \times 1$ convolutional layers, which are initialized with zeros.
 The control module highlighted in green in Figure \ref{fig:control} is a pivotal component of the proposed denoiser architecture. Following the ControlNet's structure \cite{zhang2023adding}, this module adopts the UNet encoder and middle block in the Stable Diffusion denoiser and is initialized with the pre-trained weights. The control module is comprised of four encoder blocks in addition to the middle block to handle four distinct resolutions—$64 \times 64$, $32 \times 32$, $16 \times 16$, and $8 \times 8$—capturing a wide range of feature abstraction scales. The outputs from these blocks are fed into the four decoder blocks and the middle block of the Stable Diffusion UNet. Before being fed into the UNet, the control module's outputs go through a series of $1 \times 1$ convolutional layers (`zero conv' layers in Figure \ref{fig:control}). 
 
 It is worth noting that the pre-trained Stable Diffusion UNet (in blue in Figure \ref{fig:control}) in the denoiser is adopted without retraining whereas the control module (in green) is trained for the proposed control module input. We obtain the control module input by concatenating the latent representation $z_t$ with the spatial features $f_{v}$ from the LDM encoder $E_{ldm}$ as in \cite{lin2023diffbir}. This concatenation enhances the input's representational capacity but it extends the control module network's channel depth necessitating additional parameters beyond those present in the pre-trained model. To handle this extension, the extra parameters are set to zero at initialization to maintain a neutral effect at the start of the training process.

To encode the channel CSI, we first perform a sinusoidal encoding to the channel SNR $\gamma$ under the assumption that it lies within a fixed range. Then we concatenate it with the real and imaginary part of the channel gain $h$ (characteristic of a Rayleigh Fading channel) before processing through a multi-layer perception (MLP) to produce the CSI embeddings. These embeddings share the same dimensionality with the time embeddings. In the control module, we integrate the CSI embeddings with the time embeddings by addition to inject the channel information. 
Furthermore, textual features $f_{t}$ serve as crucial prompts infusing contextual information to both the pre-trained UNet and the control module. As shown in Figure \ref{fig:control}, textual features $f_{t}$ are used as text prompts and are passed to the cross-attention layers of ViTs within each encoder/middle/decoder block as keys and values.  

Similar to the LDM loss (\ref{ldm}), our approach focuses on reducing the Mean Square Error (MSE) of the noise prediction while adhering to the condition set \(\bar{c}\) given in equation (\ref{condition}). Our loss function is thus represented by:
\begin{equation}
    \mathcal{L}_{diff} = \mathbb{E}_{z_0, \bar{c}, t, \epsilon}||\epsilon - \epsilon_{\theta}(\sqrt{\bar{\alpha}_t}z_0 + \sqrt{1-\bar{\alpha}_t}\epsilon, \bar{c}, t)||_2^2.
    \label{diff}
\end{equation}
This specially designed denoiser fine-tuned with a variety of conditioning variables is the key component in our framework to enhance the resilience of image reconstruction.

\algo

\subsubsection{Adding Intermediate Guidance} \label{guidance}
Though the proposed DiffJSCC does not inherently rely on intermediate guidance throughout the diffusion process, as articulated by Lin et al. \cite{lin2023diffbir}, diffusion models strive to carefully navigate the delicate trade-off between the realism of the generated images and their fidelity to the original content. It is worth mentioning that these models might unintentionally introduce mismatched elements that may look realistic but different from the actual target. This issue is more prevalent when there is a significant difference between the target and training domains. In scenarios where it is crucial to ensure that reconstructed images faithfully reflect the original content, we incorporate an optional guidance mechanism that leverages the initial JSCC reconstruction $\hat{x}$ to direct the intermediate latent state $\tilde{z}_0$ during the denoising process. Prior to applying guidance, the intermediate state $\tilde{z}_0$ at a particular time $t$ can be determined equation (\ref{forward}):
\begin{equation}
    \tilde{z}_0 = \frac{z_t - \sqrt{1-\bar{\alpha}_t} \epsilon_{\theta}(z_t, \bar{c}, t)}{\sqrt{\bar{\alpha}_t}} 
\end{equation}
Having estimated $\tilde{z}_0$, the subsequent latent state $z_{t-1}$ is inferred in accordance with equation (\ref{posterior}). To encourage greater correspondence between the denoised image and the initial JSCC reconstruction, $\tilde{z}_0$ is nudged closer to the spatial features $f_{spa}$, yielding a revised estimation $\hat{z}_0$:
\begin{equation}
    \hat{z}_0 = \tilde{z}_0 - \frac{\lambda}{C_lH_lW_l} (\tilde{z}_0 - f_{v}),
\end{equation}
where the variables $C_l$, $H_l$, and $W_l$ correspond to the dimensions of the latent space, and $\lambda$ is a scalar that regulates the level of alignment between $\hat{z}_0$ and $f_{v}$. The posterior distribution $q(z_{t-1}|z_{t}, \hat{z}_{0})$ subsequently determines the sampled state $z_{t-1}$ as given by equation (\ref{posterior}) substituting $z_{0}$ with $\hat{z}_{0}$. This entire process is captured in Algorithm \ref{alg:alg1}. It is worth emphasizing that the application of the intermediate guidance is optional and can be omitted by assigning a value of zero to $\lambda$, thereby maintaining the general functionality of the diffusion process without the additional guidance.

From the denoised $z_{0}$ (or $\hat{z}_{0}$), the final output image $\hat{x}_{diff}$ is generated using an LDM decoder $\hat{x}_{diff} = D_{ldm}(z_{0})$.

\section{Training and Evaluation} \label{train}

\figKodakRate
\figKodakSNR
\figKodakSNRRAY
\figVisual

\subsection{Datasets and Implementation}

\subsubsection{Dataset}

%We evaluate the proposed DiffJSCC in two distinct scenarios: general image transmission and facial image transmission. For general image transmission, DiffJSCC is trained on a subset of the OpenImage \footnote{https://storage.googleapis.com/openimages/web/index.html} dataset, comprising 300k high-quality images. During training, these images are randomly cropped to $256 \times 256$. For evaluation, the performance of DiffJSCC is assessed using the Kodak dataset, which includes 24 high-quality images of size $768 \times 512$. Furthermore, we test DiffJSCC in the ADE20K validation set for both image reconstruction and semantic segmentation tasks.
Our DiffJSCC framework evaluation encompasses two principal applications: general image transmission and specialized facial image transmission. In the general image transmission scenario, DiffJSCC is fine-tuned on a selected dataset of approximately 300,000 high-resolution images from the OpenImages collection\footnote{https://storage.googleapis.com/openimages/web/index.html}. Throughout the training phase, these images are subject to random cropping operations, resulting in dimensions of $256 \times 256$ pixels. For the performance evaluation, we use the standard Kodak dataset, which consists of 24 curated high-fidelity images with dimensions of $768 \times 512$ pixels. In addition, the performance evaluation is extended to the ADE20K validation set with 2000 images to assess image reconstruction quality and semantic segmentation accuracy.

%For facial image transmission, we train and test DiffJSCC on the CelebAHQ dataset, which contains 27,000 images for the training set and 3,000 images for the test set. The images are resized to $512 \times 512$.
In the specialized context of facial image transmission, the DiffJSCC is trained with the CelebAHQ dataset. We use a training set of 27,000 images and a separate test set of 3,000 images where all images are resized to a standardized resolution of $512 \times 512$ pixels for consistent evaluations. %This rigorous training and testing regimen ensures a thorough investigation of DiffJSCC's capabilities in handling detailed and nuanced facial imagery within contemporary transmission contexts.

\subsubsection{Implementation}

%For both the JSCC encoder and decoder, an SNR-adaptive deep JSCC architecture as described in \cite{yang2022deep} is utilized, where the SNR data is integrated with the intermediate features through the Squeeze-Excitation mechanism. The transmission rate remains constant for each model, and the policy network is excluded. The transmission rate $R$ is determined by the output channel count $C_{out}$ and the downsampling factor $D$. We set the downsampling factor to $D=4$, thus the transmission rate $R$ is given by $R=C_{out}/(2^4 \times 2^4 \times 2)=C_{out}/512$. For general image transmission, training is carried out on $256 \times 256$ random patches with a batch size of 32. For facial image transmission, training is done directly on $512 \times 512$ CelebAHQ images with a batch size of 16. The Adam optimizer is employed with an initial learning rate of $10^{-3}$. Training extends over 100,000 iterations with a cosine decay schedule.
Within the proposed framework, we utilize an SNR-adaptive JSCC architecture for both the encoder and decoder, as detailed by Yang et al.\cite{yang2022deep}. This architecture incorporates the CSI into the feature processing pipeline via a Squeeze-Excitation mechanism \cite{hu2018squeeze}, enhancing the adaptability to varying SNR conditions. The transmission rate $\rho$ is fixed for each model configuration without the policy network introduced in \cite{yang2022deep}. Within this network structure, $\rho$ is determined by the output channel count ($C_{out}$) and the downsampling factor ($D$), satisfying $\rho = C_{out}/(3 \times 2^{2D+1})$. General image transmission training involves random $256 \times 256$ patches and a batch size of 32, while the facial image dataset utilizes full $512 \times 512$ CelebAHQ images with a smaller batch size of 16. An Adam optimizer initializes learning at $10^{-3}$, tapering over 100,000 iterations following a cosine decay schedule.

%For the transmission of both general and facial images, the control module, leveraging the Stable Diffusion 2.1-base as the generative prior, is fine-tuned on $512\times512$ resized images with a batch size of 16 over 25k iterations, utilizing a learning rate of $10^{-4}$. During the training phase of the control module, the JSCC encoder and decoder are kept constant. When training both the JSCC encoder/decoder and the control module, the SNR $\gamma$ is uniformly sampled within the interval of $[0, 14]$dB for general images and $[-5, 5]$dB for facial images.
Our control module is based on Stable Diffusion 2.1-base and we adjusted it for images resized to $512 \times 512$ pixels. The batch size is 16 across 25,000 iterations at a reduced learning rate of $10^{-4}$. Note that the JSCC components remain unaltered (with fixed parameters) through the control module's training phase. The SNR, $\gamma$, is uniformly sampled within the ranges—$[0, 14]$dB for general images and $[-5, 5]$dB for facial images, ensuring adaptability to SNR fluctuations.

%All the training processes are conducted using 2 NVIDIA A40 GPUs. During inference, we use spaced DDPM sampling for 50 steps for the diffusion model. Unless otherwise stated, the scaling factor $\lambda$ is set to zero by default.
We used 2 NVIDIA A40 GPUs to accelerate training. Though the Stable Diffusion model is trained using 1000 time steps, we use a spaced DDPM sampling for 50 steps during inference adopting a common strategy \cite{rombach2022high, lin2023diffbir} to improve sampling efficiency. Unless specified otherwise, the scaling factor $\lambda$ is preset to zero as a standard baseline configuration without the intermediate guidance introduced in Section \ref{guidance}.

\subsection{Metrics}
%Our evaluation strategy employs well-established distortion metrics such as the Peak Signal-to-Noise Ratio (PSNR) and the Multiscale Structural Similarity Index (MS-SSIM) \cite{Wang2003Multiscale}. To evaluate the perceptual quality of the reconstructions, we also make use of Learned Perceptual Image Patch Similarity (LPIPS) \cite{Zhang2018Unreasonable} and Fr'echet Inception Distance (FID) scores \cite{Heusel2017GANs}. Given that the Kodak dataset comprises only 24 images, each image is transmitted five times to ensure reliable assessments. Moreover, for Kodak images, the FID score is computed using half-overlapping $256 \times 256$ image patches. To evaluate performance in the semantic segmentation task, the mIoU metric is utilized.
For evaluations, we adopt standard distortion metrics such as PSNR and MS-SSIM\cite{Wang2003Multiscale} to measure image fidelity concerning pixel intensity and structural details, respectively. To evaluate the perceptual quality of reconstructed images, we employ more recently proposed metrics such as Learned Perceptual Image Patch Similarity (LPIPS)\cite{Zhang2018Unreasonable} and Fréchet Inception Distance (FID) scores\cite{Heusel2017GANs}. These metrics provide crucial insights into visual similarity and how well the distribution of generated images aligns with that of real images. 
Given the Kodak dataset contains only 24 images, we transmit each image five times to measure our performance metrics reliably. For Kodak images, the FID score is calculated over half-overlapping patches of $256 \times 256$ pixels.
Our evaluation also includes the performance of the semantic segmentation task on the generated and original images measured by the mean Intersection over Union (mIoU) metric.

\subsection{Performance on General Image Transmission}

Figure \ref{fig:KodakRate} presents the performance comparison of our DiffJSCC framework against traditional separate source-channel coding schemes as well as the foundational DeepJSCC approach which is also used for DiffJSCC's initial image reconstruction. The evaluation is across various transmission rates while fixing the SNR to 1dB or 10dB. The benchmark separate coding scheme utilizes the BPG codec for image compression paired with LDPC codes conforming to the IEEE 802.11n WiFi standard with a specific block length. Under the 1dB SNR scenario, the LDPC code adopts a $2/3$ coding rate with BPSK modulation, while the coding rate is $1/2$ with 16-QAM modulation for the 10dB SNR case. To adjust the transmission rate, we manipulate the BPG's quantization parameter with the range from 51 to 0. The `BPG-Capacity' is the upper bound of the separate source-channel coding scheme (i.e., LDPC is replaced by ideal channel coding) where channel coding is error-free when the rate is under the channel capacity calculated by $C=\text{log}_{2}(1+\gamma)$.

DeepJSCC and DiffJSCC methods support four distinct rates: $\rho=1/384, 1/192, 1/96,$ and $1/48$. Notably, the basic DeepJSCC method surpasses traditional separate coding methods across multiple metrics, validating previous studies. Though the diffusion-based refinements from our DiffJSCC may induce degradation in traditional distortion metrics such as PSNR and MS-SSIM similar to \cite{careil2023towards, lin2023diffbir}, DiffJSCC exhibits remarkable improvements in visual quality assessed by LPIPS and FID metrics. These metrics capture the capability of DiffJSCC to produce better photorealistic reconstructions even at compression levels unachievable by the other schemes (including theoretical `BPG-Capacity').

Notice, at a 1dB SNR, DiffJSCC attains a low FID score of $117.9$ at the transmission rate of merely $\rho=1/384=0.0026$, outperforming the `BPG+Capacity' baseline's FID score of $165.4$ at a substantially higher rate of $\rho=0.0377$. This translates to DiffJSCC achieving superior image quality by a margin of $28.7\%$ while utilizing $14\times$ less symbols. In scenarios where the SNR is increased to 10dB, DiffJSCC continues to hold a significant advantage over the other methods although the performance gap in terms of LPIPS and FID is reduced. When transmitting at a rate of $\rho=0.0026$, the FID score of DiffJSCC demonstrates an impressive $64.3\%$ enhancement over DeepJSCC and a $70.7\%$ over `BPG+Capacity'. This confirms the effectiveness of the DiffJSCC framework in reconstructing highly realistic images under stringent rate and channel constraints.

In Figure \ref{fig:KodakSNR} and \ref{fig:KodakSNRRAY}, we assess DiffJSCC in comparison to other baseline approaches across various SNR conditions for AWGN channel and Rayleigh Fading channel respectively. The DeepJSCC-$adv$ model from Yang et al.\cite{yang2022ofdm} is included for comparison, which uses the hyperparameter $\alpha$ to adjust the weight of adversarial loss. Various $\alpha$ parameters were tested to explore the trade-off between distortion and visual quality. Similarly, we experimented with different $\lambda$ values in DiffJSCC with the same goal. These evaluations were conducted at transmission rates of $\rho=1/384$ and $\rho=1/96$, with SNR values set at $\gamma=1, 4, 7, 10, 13$ dB. All benchmarks including ours are based on the same deep JSCC network architecture.
The results in Figure \ref{fig:KodakSNR} demonstrate that although the DeepJSCC-$adv$ baseline successfully lowers LPIPS, it struggles to effectively minimize FID scores regardless of the $\alpha$ value used. The FID score remained above $190$ at $\rho=1/384$ and exceeded $110$ at $\rho=1/96$. Conversely, DiffJSCC significantly improves both LPIPS and FID scores over the entire range of SNR values, while maintaining similar PSNR and MS-SSIM distortion metrics. 
At $\lambda=0$ and $\rho=1/384$, DiffJSCC shows a notable average improvement of $41.4\%$ in LPIPS and $56.7\%$ in FID scores compared to the standard DeepJSCC. When compared with DeepJSCC-$adv$ using $\alpha=5\times10^{-3}$, it achieves $21.1\%$ better LPIPS and $52.2\%$ enhanced FID metrics. At $\rho=1/96$, DiffJSCC consistently demonstrates a significant gain in performance, improving average LPIPS and FID scores by $43\%$ and $53.2\%$ over DeepJSCC, respectively, and $18.9\%$ and $48.4\%$ compared to DeepJSCC-$adv$.
Increasing the scale factor $\lambda$ from 0 to 100 gradually aligns the generated image with the initial JSCC reconstruction, enhancing PSNR and MS-SSIM but reducing perceptual quality. However, this trade-off still leads to better visual perception metrics (LPIPS and FID) than the baselines. This significant improvement in the perceived visual quality highlights the effectiveness of our fine-tuning control module along with the Stable Diffusion model. A similar trend can be observed under the Rayleigh Fading channel as shown in Figure \ref{fig:KodakSNRRAY}. 

%Additionally, visualizations in Figure \ref{fig:KodakVis} illustrate the enhanced perceptual quality attained by DiffJSCC at 1dB SNR. When $R=1/128$, the DeepJSCC reconstruction exhibits significant distortion and fails to recover critical details. Conversely, DiffJSCC yields the optimal perceptual quality, effectively restoring the main structures of the lighthouse and the house, although with minor deviations from the original image. With a transmission rate increased to $1/32$, DiffJSCC continues to excel, producing content more closely aligned with the source image. Moreover, we present the captions generated by Blipv2 based on the initial JSCC reconstructions. The generated captions are shown to be resilient to the quality of these initial reconstructions, containing the appropriate keywords in the image (highlighted in red).
Figure \ref{fig:KodakVis} provides visual examples of the superior perceptual quality achieved by DiffJSCC at a challenging SNR of 1dB under the AWGN channel. At a transmission rate of $\rho=1/384$, the reconstructions produced by the conventional DeepJSCC framework exhibit significant distortion and have difficulty recovering important image details. Conversely, DiffJSCC demonstrates a remarkable capacity to restore crucial elements of the image, like the structures of the lighthouse and house, offering significantly enhanced perceptual quality despite slight variances from the original.
With the transmission rate raised to $1/96$, DiffJSCC's capabilities become more evident, resulting in reconstructions that accurately reflect the original images. Furthermore, we demonstrate the validity of captions generated by the BLIPv2 model based on the initial JSCC reconstructions. These captions prove to be robust against variations in reconstruction quality, consistently capturing the important content keywords of each image (highlighted in red). This confirms the ability of DiffJSCC to convey essential semantic information even at reduced transmission rates.

\figADE
\figVisADE

\figCelebA 
\figVisCelebA

%Next, Figure \ref{fig:figADE} illustrates the performance on the ADE20k dataset at $R=1/128$ and $R=1/32$, presenting its results in terms of FID and mIoU. For the mIoU assessment, we generate semantic segmentation masks on the reconstructed images using the pre-trained ViT-adaptor \cite{chen2022vision}. It is evident that while DeepJSCC-$adv$ can effectively lower the FID scores, it results in a significant drop in mIoU, rendering it ineffective for downstream tasks. In contrast, the proposed DiffJSCC enhances both FID and mIoU significantly. When $R=1/128$, DiffJSCC offers an average boost of $69.2\%$ in FID score and $52.5\%$ in mIoU in comparison to DeepJSCC, along with an average improvement of $75.6\%$ in FID score and $161\%$ in mIoU relative to DeepJSCC-$adv$. For $R=1/32$, it delivers an average enhancement of $42.9\%$ in FID score and $8.4\%$ in mIoU over DeepJSCC, as well as an average gain of $33.3\%$ in FID score and $19.1\%$ in mIoU over DeepJSCC-$adv$. As SNR increases, DiffJSCC's performance in mIoU closely approaches that of the undistorted original images. Additionally, we offer some visualizations on ADE20K in Figure \ref{fig:ADEVis}. For DeepJSCC, while grass and buildings are somewhat recognizable, roads and lamps are unclear. For DeepJSCC-$adv$, the pre-trained ViT-adaptor fails completely, showing large deviations from real-world images. In contrast, the image produced by DiffJSCC demonstrates superior visual quality and retains the same labels for most areas as the original undistorted image.
In Figure \ref{fig:figADE}, we assess the DiffJSCC framework using the ADE20K dataset at transmission rates $\rho=1/384$ and $\rho=1/96$ under the AWGN channel, examining FID scores and mIoU for semantic segmentation performance. For mIoU evaluations, semantic masks are inferred from the reconstructed images utilizing a pre-trained ViT-adaptor \cite{chen2022vision}. The data indicates that while the DeepJSCC-$adv$ approach can effectively reduce FID scores, it comes at the cost of decreased mIoU, thus limiting its practicality for subsequent image processing tasks.
In contrast, DiffJSCC notably improves both FID and mIoU. When $\rho=1/384$, it demonstrates an average improvement (reduction) in FID by $69.2\%$ and a $52.5\%$ increase in mIoU over DeepJSCC, alongside the notable improvement of $75.6\%$ in FID and $161\%$ in mIoU compared to DeepJSCC-$adv$. For the higher transmission rate $\rho=1/96$, DiffJSCC consistently shows superior performance, with a mean enhancement of $42.9\%$ in FID and $8.4\%$ in mIoU over DeepJSCC, along with gains of $33.3\%$ in FID and $19.1\%$ in mIoU in comparison to DeepJSCC-$adv$. With increasing SNR values, the mIoU obtained by DiffJSCC begins to align closely with that of the original undistorted images.
Figure \ref{fig:ADEVis} presents a visual example from the ADE20K dataset. The example shows that, with DeepJSCC, elements such as grass and buildings are somewhat recognizable, yet roads and lamps appear unclear. The reconstructions obtained from DeepJSCC-$adv$ significantly lack authenticity, and the ViT-adaptor fails to generate accurate segmentation masks. On the other hand, images produced by DiffJSCC demonstrate higher visual quality to produce more accurate segmentation labels in most regions, thus highlighting the framework's applicability for practical JSCC scenarios.

\subsection{Performance on Facial Image Transmission}

Figure \ref{fig:CelebASNR} presents a comparison of our DiffJSCC framework with DeepJSCC and other generative JSCC methods, including InverseJSCC \cite{erdemir2023generative} and the approach presented by Yilmaz et al. \cite{yilmaz2023high} on CelebAHQ dataset. For DiffJSCC, we showcase results with varying $\lambda$ values of 0, 10, and 50. The outcomes of InverseJSCC and Yilmaz et al. are taken from their original papers. We only evaluate the PSNR and LPIPS metrics since MS-SSIM and FID evaluations were not conducted in their works.
Overall, DiffJSCC shows superior performance, surpassing Inverse-JSCC regarding both PSNR and LPIPS metrics. In comparison with the Yilmaz et al. method, DiffJSCC has a slightly lower PSNR but achieves superior LPIPS scores. Unlike the Kodak image results, where $\lambda=0$ provided the optimal LPIPS, this configuration introduces artifacts because of the significant domain gap between the Stable Diffusion training dataset and the CelebAHQ dataset. However, by utilizing direct guidance with the initial JSCC reconstructions (with $\lambda=50$), these artifacts are effectively mitigated, resulting in enhanced LPIPS performance.
Quantitatively, at $\lambda=50$, DiffJSCC accomplishes an average enhancement of $15.9\%$ and $28.6\%$ over InverseJSCC for $\rho=1/768$ and $\rho=1/192$, respectively, and showcases an improvement of $10.3\%$ and $9.4\%$ for $\rho=1/768$ and $\rho=1/192$ against the Yilmaz et al. method. Figure \ref{fig:CelebAVis} visualizes examples from the CelebA dataset, highlighting DiffJSCC's ability to reconstruct high-quality images in extremely challenging conditions such as a very low bit rate ($\rho=1/768$) and a low SNR value ($\text{SNR}=-5\text{dB}$). As the SNR increases, the quality of the reconstructions improves, making them more closely resemble the original images. %It is noteworthy that, at $\lambda=0$, unintended artifacts in the DiffJSCC reconstructions are more noticeable, especially at $-5$dB SNR. However, when direct guidance with $\lambda=50$ is applied, these artifacts are reduced, resulting in higher visual quality.

\subsection{Ablation Studies}

\subsubsection{The effect of different conditions}

%In this section, we analyze the effect of different conditional signals on the performance of the control module, as illustrated in Figure \ref{fig:ablation} bottom. It is evident that depending solely on spatial features from the initial JSCC reconstruction leads to less optimal outcomes. Adding channel SNR as a supplementary condition noticeably improves FID scores by $2\%$ on average, especially at lower SNR. This indicates that including channel state information such as SNR in the denoising model helps it adapt to various distortion scenarios with different SNRs. Moreover, introducing captions as textual prompts to the control module further enhances the performance of the FID score by $4.3\%$. As depicted in Figure \ref{fig:ablation} top, the quality of generated captions, assessed by the CLIP score \cite{radford2021learning}, decreases as the SNR and transmission rate decrease, yet the system's performance improvement remains significant at low SNRs as it retains the keywords in the image, demonstrating the robustness and effectiveness of the textual information. The effectiveness of the textual modality is also evident in the additional improvement when captions generated from undistorted images are used, as shown by the orange line in Figure \ref{fig:ablation} below.
In this analysis, we investigate the impact of various conditional signals on the control module's performance on the Kodak dataset under the AWGN channel, as shown in Figure \ref{fig:ablation} right, emphasizing the importance of including multimodal information during the image restoration process. We observe that relying exclusively on spatial features extracted from the initial JSCC reconstruction offers limited effectiveness, whereas augmenting the model with channel SNR information yields an average FID score improvement of $2\%$, a benefit that becomes more pronounced at reduced SNR levels. %This enhancement underlines the critical role of channel state information in enabling the denoiser to adjust to varying levels of distortion. 
Incorporating captions as textual prompts into the denoising framework further elevates FID performance by an additional $4.3\%$. Figure \ref{fig:ablation} left showcases that despite a natural decline in caption quality measured by the CLIP score\cite{radford2021learning} as SNR and transmission rates decrease, the presence of image keywords ensures sustained improvements in FID as shown in Figure \ref{fig:ablation} right. This outcome illustrates the resilience and utility of textual cues in the denoising process. The benefit is further demonstrated by the marked performance increase when using captions derived from undistorted images, as indicated by the orange line in Figure \ref{fig:ablation} right.

\figAblation
\figAblationVis
\figStep

Figure \ref{fig:ablationVis} serves as a visualization example of the influence of textual cues: Despite the original JSCC reconstruction being affected by blur and distortion, The BLIPv2 model employed in our framework can generate coherent text prompts that capture the main context of the image. Without textual cues (Fig. \ref{fig:ablationVis} bottom left), our approach can improve the general image quality but struggles to accurately recreate the parrot on the left. Incorporating text prompts addresses this issue significantly (Fig. \ref{fig:ablationVis} bottom right), enabling the realistic restoration of intricate details like the parrot's head, demonstrating the crucial importance of textual information in the image reconstruction process.

%To better demonstrate the impact of textual cues, we present a representative example in Figure \ref{fig:ablationVis}. Although the initial JSCC reconstruction is blurred and distorted, the BLIPv2 model successfully extracts coherent text prompts that capture the image's core semantic content. Without textual cues, our method can enhance visual clarity, but fails to properly restore the parrot on the left side of the image. However, when text prompts are incorporated, the left parrot's reconstruction significantly improves, with finer details such as the bird's head being accurately recovered. 

\subsubsection{The effect of the number of sampling steps}

%We analyze the impact of changing the number of sampling steps in the Kodak dataset for LPIPS and FID scores with $R=1/128$ across multiple SNR values, as presented in Figure \ref{fig:ablationSteps}. An ongoing improvement in both LPIPS and FID metrics is noticed as the sampling steps increase. Importantly, the metrics plateau around 50 steps for all SNR values. Taking $\text{SNR}=13\text{dB}$ as an example, the improvements in the LPIPS and FID scores from 25 steps to 50 steps are $0.02$ and $6.5$, while the values become $0.004$ and $2.5$ from 50 steps to 100 steps. Thus, results justify the use of 50 sampling steps in all our experiments and strike a better balance between inference time and performance.
This section investigates the impact of the number of denoising sampling steps on LPIPS and FID scores using the Kodak dataset with a transmission rate of $\rho=1/384$. The performance evolution across different SNR levels is illustrated in Figure \ref{fig:ablationSteps}. It is evident that both LPIPS and FID metrics enhance considerably as the number of sampling steps increases. Notably, after reaching 50 steps, these enhancements level off with marginalized gains from additional steps across all evaluated SNR values.
When $\text{SNR}=13\text{dB}$, for example, increasing the sampling from 25 to 50 steps results in substantial gains of $0.02$ for LPIPS and $6.5$ for FID, whereas increasing it from 50 to 100 steps yields reduced improvements of $0.004$ for LPIPS and $2.5$ for FID. These findings led to our decision to use 50 sampling steps in our experiments in earlier sections. Using 50 steps balances the efficiency and effectiveness of our method considering the required computational resources, inference time, and reconstruction image quality.

\section{Conclusion}

In this study, we introduce DiffJSCC, a novel deep JSCC framework that capitalizes on the robust priors of the pre-trained Stable Diffusion model to substantially improve the perceptual quality of reconstructed images. Unlike the traditional end-to-end learning approach, DiffJSCC initially derives multimodal conditions from the received symbols. For enhanced training stability and efficiency, DiffJSCC employs a JSCC decoder to create an initial reconstruction of the transmitted image, which is then utilized to generate multimodal features. DiffJSCC effectively integrates multimodal conditions that consist of spatial, textual, and channel state information to steer the image generation process. Our framework employs a new control module to fine-tune the Stable Diffusion model based on the generated conditions, and it leverages the initial JSCC reconstruction to guide the denoising diffusion process, balancing fidelity and realism. Extensive experiments on both general and facial image transmission cases demonstrate that DiffJSCC considerably outperforms the latest generative JSCC methods, providing superior human-perceptual image quality and downstream task performance, particularly in environments with low transmission rates and poor SNRs. The proposed DiffJSCC is model-agnostic and can be extended with more advanced deep JSCC architectures and/or improved conditional diffusion models. Furthermore, DiffJSCC introduces no extra computations on the transmitter side, which is particularly beneficial for scenarios with energy-limited image-capturing devices.

\bibliographystyle{IEEEtran}
\bibliography{refs}

\end{document}